\def\paperversion{2} % 1 - conference version, 2 - full version

\documentclass[conference,twocolumn,romanappendices]{IEEEtran}

\IEEEoverridecommandlockouts

\usepackage{etex}

\usepackage[subtle,indent=normal,mathspacing=normal]{savetrees}

\usepackage[dvipsnames]{xcolor}

\usepackage{balance}
\usepackage{todonotes}

\usepackage{cite}
\usepackage{algorithmic}
\usepackage{graphicx}
\usepackage{textcomp}

\usepackage[cmex10]{amsmath}
\usepackage{amssymb}
\usepackage{amsthm}
\usepackage{amsfonts}
\usepackage{mathtools}
\usepackage{stmaryrd}
\usepackage{url}
\usepackage{enumerate}
\usepackage{float}
\usepackage{booktabs}
\usepackage{multirow}
\usepackage{colortbl}
\usepackage[font=footnotesize]{caption}
\usepackage[font=footnotesize]{subcaption}

\usepackage[T1]{fontenc}
\usepackage[capitalise]{cleveref}
\usepackage{pifont}
\usepackage[utf8]{inputenc}
\usepackage[greek,english]{babel}
\usepackage{CJKutf8}
\addto\captionsenglish{}

\usepackage{tikz}
\usepackage{tikz-qtree}
\usepackage{tikz-3dplot}
\tdplotsetmaincoords{70}{20}
\usetikzlibrary{backgrounds,fit}
\usetikzlibrary{matrix,calc,arrows}
\usetikzlibrary{shapes.geometric}

\usepackage{pgfplots}

\newtheorem{lemma}{Lemma}
\newtheorem{theorem}{Theorem}

\theoremstyle{definition}\newtheorem{definition}{Definition}
\theoremstyle{remark}\newtheorem{example}{Example}
\theoremstyle{remark}

\newcommand{\ellmax}{\ell^\text{max}}
\newcommand{\core}{C^{\odot}}
\newcommand{\wc}{C^{\S}}
\newcommand{\ccore}{\mathbf{c}^{\odot}}
\newcommand{\overlap}{\supset\mkern-7mu\subset}
\newcommand{\prefix}{\not\mkern-3.5mu\overlap}
\newcommand{\ZZ}{\mathbb{Z}}
\newcommand{\NN}{\mathbb{N}}

\DeclareMathOperator{\len}{len}
\DeclareMathOperator{\vol}{Vol}

\begin{document}

%\title{On Entropy-Achieving Multi-Channel Prefix-Free Codes That Cannot Be Tree-Decodable}
%\title{Existence of Optimal Multichannel Tree-Decodable Codes That Are Not Optimal Prefix Codes}
\title{Multichannel Optimal Tree-Decodable Codes\\are Not Always Optimal Prefix Codes}

\author{
	\IEEEauthorblockN{
		Hoover~H.~F.~Yin,
		Harry~W.~H.~Wong,
		Mehrdad~Tahernia, and
		Russell~W.~F.~Lai
	}
	\thanks{
		H.~H.~F.~Yin is with the Institute of Network Coding, The Chinese University of Hong Kong.
		H.~W.~H.~Wong is with the Department of Information Engineering, The Chinese University of Hong Kong.
		M.~Tahernia is an independent researcher. %with the Centre for Perceptual and Interactive Intelligence (CPII) Ltd.
		%R.~W.~F.~Lai is with the Chair of Applied Cryptography, Friedrich-Alexander University Erlangen-Nuremberg.
		R.~W.~F.~Lai is with the Chair of Applied Cryptography, Friedrich-Alexander-Universit\"at Erlangen-N\"urnberg.
\ifnum\paperversion=1
Omitted proofs of this paper can be found in \cite{prefix2}.
\fi
	}
	%\IEEEauthorblockN{
	%	Hoover~H.~F.~Yin\IEEEauthorrefmark{1},
	%	Harry~W.~H.~Wong\IEEEauthorrefmark{2}, and
	%	Mehrdad~Tahernia\IEEEauthorrefmark{3}
	%	%Russell~W.~F.~Lai\IEEEauthorrefmark{4}
	%}
	%\IEEEauthorblockA{\IEEEauthorrefmark{1}Institute of Network Coding, The Chinese University of Hong Kong}
	%\IEEEauthorblockA{\IEEEauthorrefmark{2}Department of Information Engineering, The Chinese University of Hong Kong}
	%\IEEEauthorblockA{\IEEEauthorrefmark{3}Centre for Perceptual and Interactive Intelligence (CPII) Limited}
	%%\IEEEauthorblockA{\IEEEauthorrefmark{4}TODO}
}

\maketitle

\begin{abstract}
	The theory of multichannel prefix codes aims to generalize the classical theory of prefix codes.
	Although single- and two-channel prefix codes always have decoding trees, the same cannot be said when there are more than two channels.
	One question is of theoretical interest: Do there exist optimal tree-decodable codes that are not optimal prefix codes?
	Existing literature, which focused on generalizing single-channel results, covered little about non-tree-decodable prefix codes since they have no single-channel counterparts.
	In this work, we study the fundamental reason behind the non-tree-decodability of prefix codes.
	By investigating the simplest non-tree-decodable structure, we obtain a general sufficient condition on the channel alphabets for the existence of optimal tree-decodable codes that are not optimal prefix codes.
\end{abstract}

\section{Introduction}

Prefix-free codes, or conventionally called \emph{prefix codes}, are a class of zero-error uniquely decodable source code being applied in a wide range of scenarios including the country codes \cite{itu}, UTF-8 \cite{utf8,helloworld}, and most importantly, data compression \cite{shannon,fano,huffman}.
In a traditional sense, an optimal prefix code is a symbol-by-symbol prefix code having the lowest redundancy when the probability of the information source is known.\footnote{%
	It is possible for a non-symbol-by-symbol prefix code to achieve a lower redundancy, e.g., the arithmetic codes \cite{ac1,ac2}.
	It is also possible for a symbol-by-symbol non-instantaneous code to do so, e.g., the AIFV codes \cite{aifv,aifv2}.
}
Huffman code \cite{huffman} is an iconic optimal prefix code which can be encoded in linear time (in the support size) for sorted probability \cite{linear_huffman} and be decoded in linear time (in the codeword length) by using the constructed decoding tree.
%This shows the existence together with the practical application of optimal prefix codes for finite source alphabets.
To better understand the nature of prefix codes, literature also include theoretical research such as those for infinite sources \cite{linder97,chow98,klimesh06,klimesh08}.

Another theoretical generalization of prefix codes is to use more channels \cite{yao10}.\ifnum\paperversion=2
\footnote{%
	Multichannel prefix codes can be interpreted as multidirectional context sets \cite{context}.
	Although some structural results in context sets are isomorphic to those in prefix codes, e.g., context trees versus decoding trees, the optimization problems on the relevant structures are quite different.
}
\fi
Not all single-channel results hold when generalized to the multichannel setting.
One example is that the satisfiability of the multichannel Kraft inequality \cite{yao10} is insufficient for the existence of prefix codes.
Instead, a rectangle packing formulation is needed to capture the geometry of prefix codes \cite{packing,packing2}.
Worse, \cite{yao10} showed an example of a prefix code having no decoding tree.
While single- and two-channel prefix codes are tree-decodable \cite{prefix}, for three or more channels, we are only assured that tree-decodable codes are prefix codes \cite{yao10}, but not the converse.
Non-tree-decodable prefix codes are not well-studied, to say the least, as they have no single-channel counterparts.

When the channel alphabet sizes are the same, an optimal tree-decodable code is an optimal prefix code \cite{yao10}, which can be constructed by manipulating the decoding tree of the single-channel Huffman code.
Much less is known for differing alphabet sizes.
To begin, although a modified Huffman procedure can produce an optimal multichannel tree-decodable code \cite{prefix}, the procedure is not known to be efficiently computable at this moment.
Furthermore, it is unclear whether the resulting tree-decodable code is also optimal as a prefix code.
One question is of interest: is an optimal tree-decodable code also an optimal prefix code in general?

In this work, we first study the fundamental reason -- the \emph{interweave structures} -- behind the non-tree-decodability of prefix codes.
After that, we make use of the simplest interweave structure to construct a class of non-tree-decodable prefix codes called the \emph{selvage codes}.
By investigating the selvage codes, we prove that if the channel alphabets %$Q$
satisfy a ``separable'' property, then there exists an optimal tree-decodable code which is not an optimal prefix code. %, or ``$Q$ is above tree line'' for short.
%satisfy a property which we called ``$t$-separable'' for $t \geq 3$, then there exists an optimal tree-decodable code which is not an optimal prefix code. %, or ``$Q$ is above tree line'' for short.

\section{Preliminaries}
%\section{Multichannel Prefix Codes}

Denote by $\NN$ and $\ZZ^+$ the sets of non-negative integers and positive integers respectively.
For any $q \in \ZZ^+$, define $\ZZ_q = \{0, 1, \ldots, q-1\}$.
% and $\ZZ_q^+ = \{1, 2, \ldots, q\}$.
We always count objects from the 0-th.
%
%Let $\mathbb{P}$ be the set of all positive primes.
The notation $\uplus$ denotes the multiset sum. %\footnote{%
%	A multiset is a set which allows multiple instances for each of its elements.
%	The multiset sum is a multiset version of union which keeps all elements of the operands.
%	For example, let $A = \{1, 2\}$ and $B = \{2, 3\}$, then $A \uplus B = \{1, 2, 2, 3\}$. %while $A \cup B = \{1, 2, 3\}$.
%}
%Denote by $e$ the Euler's number.
Let $\epsilon$ be the empty string.
%For any strings $a, b$, denote by $a \| b$ the concatenation of $a$ and $b$. %, e.g., $01 \| 10$ means the string $0110$.
%To represent a string formed by duplicating the same symbol, we bold the symbol and write the number of repetitions as its exponent, e.g., $\mathbf{0}^4$ means the string $0000$.
%For any $s \in \NN$, we can write its $q$-ary number representation uniquely as a string of length $\lceil \log_q (s+1) \rceil$.
%Let $\Lbag s \Rbag_{q}^{\ell}$ be the string of the $q$-ary number representation of $s$ left-padded with $0$s such that the length becomes $\ell$.
%Further, define $\Lbag w \Rbag_{q}^{-1}$ to be the unique $s \in \NN$ such that $w$ is the $q$-ary number representation of $s$.
%For example, $\Lbag 7 \Rbag_2^4$ is the string $0111$, and $\Lbag 0111 \Rbag_2^{-1}$ gives back $7$.

\subsection{Multichannel Prefix Codes}

Suppose there are $n$ channels.
For each $i \in \ZZ_n$, each symbol sent on the $i$-th channel is from the alphabet $\mathcal{Z}_i$ where $|\mathcal{Z}_i| = q_i \ge 2$.
Since we can map $\mathcal{Z}_i$ to $\ZZ_{q_i}$ bijectively, we assume $\mathcal{Z}_i = \ZZ_{q_i}$ in the rest of this work.
For any $k \in \ZZ^+$, we write $\ZZ_{q_i}^k$ as the set of strings of $k$ symbols from $\ZZ_{q_i}$.
Define $\ZZ_{q_i}^0 = \{\epsilon\}$.

For any $n \in \ZZ^+$, define $Q_n \coloneqq (q_0, q_1, \ldots, q_{n-1}) = (q_i)_{i \in \ZZ_n}$.
We drop the subscript of $Q_n$ when it is clear from context.
A $Q_n$-ary \emph{word} is an $n$-tuple where the $i$-th component of the word is in $\mathcal{Z}_i^\ast \coloneqq \bigcup_{k = 0}^\infty \ZZ_{q_i}^k$, the set of all possible strings built using the alphabet $\ZZ_{q_i}$.
For any word $\mathbf{c}$, the $i$-th component is denoted by $\mathbf{c}(i)$.
%For simplicity, we define the concatenation of two codewords $\mathbf{c}, \mathbf{c}'$, denote by $\mathbf{c} \| \mathbf{c}'$, be a word where its $i$-th component is $\mathbf{c}(i) \| \mathbf{c}'(i)$.

Let $Z$ be an information source and $\mathcal{Z}$ be the alphabet of $Z$.
A \emph{$Q_n$-ary source code} for $Z$ is a map $\mathcal{Q} \colon \mathcal{Z} \to \prod_{i \in \ZZ_n} \mathcal{Z}_i^\ast$.
For each $z \in \mathcal{Z}$, $\mathcal{Q}(z)$ is the \emph{codeword} for $z$.
The $i$-th component of any codeword is sent through the $i$-th channel.
When we send more than one codeword, the codewords are concatenated component-wise so that the boundaries of the codewords are not explicit.
The image of $\mathcal{Q}$ is called the \emph{codebook} of the source code.
For convenience, we also refer a source code to its codebook.
The \emph{codeword matrix} $M$ of a $Q_n$-ary source code $C = \{\mathbf{c}_j\}_{j \in \ZZ_m}$ is an $m \times n$ matrix where the $j$-th row of $M$ is $\mathbf{c}_j$.
If the multiset $P$ is the probability of $Z$, then the source code for $Z$ is also called a source code on $P$.

%For example, if we want to send the following three codewords $(11,0)$, $(1, 1)$ and $(0, \epsilon)$ sequentially, then what we actually send is $(1110, 01)$.

Two words $\mathbf{c}, \mathbf{c}'$ are \emph{prefix-free}, denoted by $\mathbf{c} \prefix \mathbf{c}'$, if there exists a channel $i$ such that $\mathbf{c}(i)$ and $\mathbf{c}'(i)$ are prefix-free.
%The notation $\prefix$ symbolizes ``no overlapping'', which will be discussed later in \cref{sec:rpg}.

\begin{definition}[Prefix Codes~\cite{yao10}]
	A \emph{$Q$-ary prefix code} is a $Q$-ary source code such that every pair of codewords are prefix-free.
\end{definition}

\begin{figure*}
	\begin{minipage}[t]{.955\textwidth}
	\centering
	\begin{minipage}[t]{.3\textwidth}
	\vskip -1em
\begin{figure}[H]
	\centering
	\scriptsize
	\tdplotsetmaincoords{70}{10}
	\begin{tikzpicture}[scale=.9]
		\begin{scope}[tdplot_main_coords]
			\coordinate (r0) at (0, 0, 0);
			\coordinate (r1) at (0, 2, 0);
			\coordinate (r2) at (2, 2, 0);
			\coordinate (r3) at (2, 0, 0);
			\coordinate (b0) at (2, 0, 0);
			\coordinate (b1) at (4, 0, 0);
			\coordinate (b2) at (4, 1, 0);
			\coordinate (b3) at (2, 1, 0);
			\coordinate (g0) at (3, 1, 0);
			\coordinate (g1) at (4, 1, 0);
			\coordinate (g2) at (4, 2, 0);
			\coordinate (g3) at (3, 2, 0);
			\foreach \x in {0, 1, 2, 3, 4}{
				\draw (\x, 0, 0) -- ++(0, 2, 0);
			}
			\foreach \y in {0, 1, 2}{
				\draw (0, \y, 0) -- ++(4, 0, 0);
			}
			\fill[red,opacity=.3] (r0) -- (r1) -- (r2) -- (r3) -- cycle;
			\fill[blue,opacity=.3] (b0) -- (b1) -- (b2) -- (b3) -- cycle;
			\fill[green,opacity=.3] (g0) -- (g1) -- (g2) -- (g3) -- cycle;
			\draw[ultra thick,red] (r2) -- (r3);
			\draw[ultra thick,blue] (b2) -- (b3);
			\draw[ultra thick,OliveGreen] (g0) -- (g3);
		\end{scope}

		\node[draw,circle,inner sep=.1em] (root) at (2.2,2.5) {$0$};
		\node (r) at (1.2,2) {};
		\node[draw,circle,inner sep=.1em] (c1) at (3.2, 2) {$1$};
		\node (b) at (2.5, 1.5) {};
		\node[draw,circle,inner sep=.1em] (c1d1) at (3.9, 1.5) {$0$};
		\node (g) at (4.4,1) {};
		\foreach \i in {0, 1, 2, 3}{
			\draw[red,opacity=.3] (r\i) -- (r.center);
			\draw[blue,opacity=.3] (b\i) -- (b.center);
			\draw[OliveGreen,opacity=.3] (g\i) -- (g.center);
		}
		\draw[ultra thick,red] (root) -- node[black,above] {$0$} (r.center);
		\draw[ultra thick,red] (root) -- node[black,above] {$1$} (c1);
		\draw[ultra thick,blue] (c1) -- node[black,above] {$0$} (b.center);
		\draw[ultra thick,blue] (c1) -- node[black,above] {$1$} (c1d1);
		\draw[ultra thick,OliveGreen] (c1d1) -- node[black,above] {$1$} (g.center);
		\filldraw (r.center) circle (2pt);
		\filldraw (b.center) circle (2pt);
		\filldraw (g.center) circle (2pt);
	\end{tikzpicture}
	\caption{An example of the relation between guillotine-cuts and RPG.}
	\label{fig:2dtree}
\end{figure}
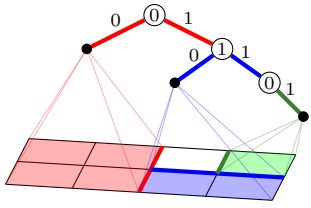
	\end{minipage}~\vrule~
	\begin{minipage}[t]{.7\textwidth}
	\vskip -1em
\begin{figure}[H]
	\centering
	\tiny
	\begin{subfigure}{.35\textwidth}
		\centering
		\tdplotsetmaincoords{82}{40}
		\begin{tikzpicture}[scale=.6]
			\foreach \z in {0, 1, 2}{
				\begin{scope}[tdplot_main_coords,canvas is xy plane at z=\z]
					\foreach \r in {1, 2, 3}
					\draw (0,0) circle (\r);
				\end{scope}
			}
			\foreach \x in {1, 2, 3}{
				\begin{scope}[tdplot_main_coords]
					\draw (\x,0,0) -- ++(0,0,2);
					\draw (-\x,0,0) -- ++(0,0,2);
					\draw (0,\x,0) -- ++(0,0,2);
					\draw (0,-\x,0) -- ++(0,0,2);
				\end{scope}
			}
			\foreach \z in {0, 1, 2}{
				\begin{scope}[tdplot_main_coords]
					\draw (1,0,\z) -- ++(2,0,0);
					\draw (-1,0,\z) -- ++(-2,0,0);
					\draw (0,1,\z) -- ++(0,2,0);
					\draw (0,-1,\z) -- ++(0,-2,0);
				\end{scope}
			}
			\foreach \i in {1, 2, 3}{
				\pgfmathsetmacro{\xx}{\i*cos(40)}
				\pgfmathsetmacro{\yy}{\i*sin(40)}
				\draw[dashed,tdplot_main_coords] (\xx,\yy,0) -- ++(0,0,2);
				\draw[dashed,tdplot_main_coords] (-\xx,-\yy,0) -- ++(0,0,2);
			}
			\begin{scope}[tdplot_main_coords]
				\pgfmathsetmacro{\xx}{3*cos(40)}
				\pgfmathsetmacro{\yy}{3*sin(40)}
				\coordinate (z0) at (\xx,\yy,0);
				\coordinate (z1) at (\xx,\yy,2);
			\end{scope}

			\fill[tdplot_main_coords,blue,opacity=.3] (-3,0,0) -- ++(2,0,0) -- ++(0,0,2) -- ++(-2,0,0) -- cycle;
			\fill[tdplot_main_coords,red,opacity=.3] (0,3,0) -- ++(0,-2,0) -- ++(0,0,2) -- ++(0,2,0) -- cycle;
			\fill[tdplot_main_coords,red,opacity=.3] (0,-3,0) -- ++(0,2,0) -- ++(0,0,2) -- ++(0,-2,0) -- cycle;
			\fill[tdplot_main_coords,blue,opacity=.3] (3,0,0) -- ++(-2,0,0) -- ++(0,0,2) -- ++(2,0,0) -- cycle;

			\draw[-stealth] ($(z0)+(.2,0)$) -- node[midway,sloped,below] {2-nd channel} ($(z1)+(.2,0)$);
		\end{tikzpicture}
		\vskip -.6em
		\caption{The cylinder structure after merging the 0-th and the 1-st channels.}
		\label{fig:3d2}
	\end{subfigure}~
	\begin{subfigure}{.33\textwidth}
		\centering
		\begin{tikzpicture}[scale=.3]
			\foreach \r in {1, 2, 3}
			\draw (0,0) circle (\r);
			\draw[-stealth] (1,0) -- node[midway,sloped,below] {3-rd channel} (3,0);
			\draw[-stealth] (-1,0) -- node[midway,sloped,below] {3-rd channel} (-3,0);
			\node at (1.5,0) [above] {$0$};
			\node at (2.5,0) [above] {$1$};
			\node at (-1.5,0) [above] {$0$};
			\node at (-2.5,0) [above] {$1$};
			\draw[ultra thick,red,opacity=.3] (45:1) -- (45:3);
			\draw[ultra thick,red,opacity=.3] (-135:1) -- (-135:3);
			\draw[ultra thick,red,opacity=.3,dotted] (45:1) -- (-135:1);
			\draw[ultra thick,blue,opacity=.3] (135:1) -- (135:3);
			\draw[ultra thick,blue,opacity=.3] (-45:1) -- (-45:3);
			\draw[ultra thick,blue,opacity=.3,dotted] (-45:1) -- (135:1);
			\node at (145:3.1) [left] {$00$};
			\node at (125:3.1) [above] {$01$};
			\node at (-35:3.1) [right] {$11$};
			\node at (-55:3.1) [below] {$10$};
			\draw (225:3.7) arc (225:180:3.7) node[above,rotate=90,every text node part/.style={align=center}] {0-th and 1-st\\channels};
			\draw[-stealth] (180:3.7) arc (180:160:3.7);
		\end{tikzpicture}
		\vskip -.6em
		\caption{The top view of the cylinder structure.}
	\end{subfigure}~%\\[-1.5em]
	\begin{subfigure}{.35\textwidth}
		\centering
		\begin{tikzpicture}[scale=.6]
			\tdplotsetmaincoords{70}{30}
			\begin{scope}[tdplot_main_coords]
				\foreach \x in {0,1,2,3,4}
				\foreach \y in {0,1,2}
				\foreach \z in {0,1,2}
				{
					\draw (\x,0,\z) -- (\x,2,\z);
					\draw (0,\y,\z) -- (4,\y,\z);
					\draw (\x,\y,0) -- (\x,\y,2);
				}
				\fill[red,opacity=.3] (0,0,0) -- ++(0,2,0) -- ++(0,0,2) -- ++(0,-2,0) -- cycle;
				\fill[blue,opacity=.3] (1,0,0) -- ++(0,2,0) -- ++(0,0,2) -- ++(0,-2,0) -- cycle;
				\fill[red,opacity=.3] (2,0,0) -- ++(0,2,0) -- ++(0,0,2) -- ++(0,-2,0) -- cycle;
				\fill[blue,opacity=.3] (3,0,0) -- ++(0,2,0) -- ++(0,0,2) -- ++(0,-2,0) -- cycle;
				\fill[red,opacity=.3] (4,0,0) -- ++(0,2,0) -- ++(0,0,2) -- ++(0,-2,0) -- cycle;

				\coordinate (x0) at (0,0,0);
				\coordinate (y0) at (4,0,0);
				\coordinate (z0) at (4,2,0);
				\coordinate (z1) at (4,2,2);

				\coordinate (x1) at (.5,2,2);
				\coordinate (x2) at (1.5,2,2);
				\coordinate (x3) at (2.5,2,2);
				\coordinate (x4) at (3.5,2,2);
			\end{scope}
			\draw[-stealth] ($(x0)+(0,-.1)$) -- node[midway,sloped,below] {0-th and 1-st channels} ($(y0)+(0,-.1)$);
			\draw[-stealth] ($(y0)+(0,-.1)$) -- node[midway,sloped,below] {2-nd channel} ($(z0)+(.1,0)$);
			\draw[-stealth] ($(z0)+(.1,0)$) -- node[midway,sloped,below] {3-rd channel} ($(z1)+(.1,0)$);
			\node at ($(x1)+(0,.2)$) {$00$};
			\node at ($(x2)+(0,.2)$) {$01$};
			\node at ($(x3)+(0,.2)$) {$11$};
			\node at ($(x4)+(0,.2)$) {$10$};
		\end{tikzpicture}
		\vskip -1em
		\caption{The rectangle representation after cutting the red plane between $00$ and $10$.}
		\label{fig:3d2a}
	\end{subfigure}
	\vskip -.8em
	\caption{The representation of a $4$-D RPG by a $3$-D rectangle.}
	\label{fig:4drpg}
	%\vskip -3.2em
\end{figure}
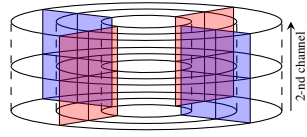
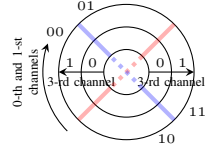
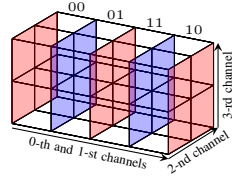
	\end{minipage}
	\end{minipage}
	\ifnum\paperversion=1
	\vskip -1.3em
	\fi
\end{figure*}

A (multichannel) decoding tree is a tree that every non-leaf node belongs to a class.
A class $i$ node means that this node is associated with the $i$-th channel.
Each branch of a class $i$ node corresponds to a distinct symbol in $\ZZ_{q_i}$, i.e., a class $i$ node has at most $q_i$ children.
Every leaf corresponds to a codeword and associates with a source symbol. %\footnote{%
	To decode a codeword, we traverse the tree from the root node.
	When we reach a class $i$ node, we read a symbol of the codeword from the $i$-th channel and traverse through the corresponding branch.
	We can decode the codeword once we reach a leaf.
%}
An example of decoding tree can be found in \cref{fig:2dtree}, where the codewords are %those used in \cref{eg:2drpg}.
$(0,\epsilon)$, $(1,0)$ and $(11,1)$.

\begin{definition}[Tree-Decodable Code~\cite{yao10}]
	A source code is a \emph{tree-decodable code} if it has a decoding tree %\footnote{%
%A decoding tree is a tree that every non-leaf node belongs to a class.
%A class $i$ node means that this node is associated with the $i$-th channel.
%Each branch of a class $i$ node corresponds to a distinct symbol in $\ZZ_{q_i}$, i.e., a class $i$ node has at most $q_i$ children.
%Every leaf corresponds to a codeword and associates with a source symbol.
%To decode a codeword, we traverse the tree from the root node.
%When we reach a class $i$ node, we read a symbol of the codeword from the $i$-th channel and traverse through the corresponding branch.
%We can decode the codeword once we reach a leaf.
%}
	such that every codeword can be decoded by this tree.
	%\footnote{%
		%Every tree-decodable code is a prefix code \cite{yao10}.
		%%Non-tree-decodable prefix codes can be decoded by the self-punctuating decoding process (SPDP) proposed in \cite{yao10}.
		%%Although in theory, prefix codes are instantaneous codes, SPDP may need to refer to the symbols of future codewords.
	%}
	A codeword matrix $M$ of a source code $C$ is called tree decodable if and only if $C$ is tree decodable.
\end{definition}

\noindent
Every tree-decodable code is a prefix code but not the converse \cite{yao10}.
All single- and two-channel prefix codes are tree decodable \cite{prefix}.

%\subsection{Entropy Bound}

%As the alphabets of the channels can be different, we need to use a unified unit to measure the amount of information.
%In this work, we use the unit ``nat'' (natural unit of information), i.e., we use $\ln$, the natural logarithm, in the evaluation of entropy.
%All results in this work are valid if we use another base for the logarithm.

\begin{figure}
		\centering
		\scriptsize
		\begin{tikzpicture}
			\node (ht) at (0,0.5) {Huffman};
			\node (tt) at (0, 0) {tree-decodable};
			\node (pt) at (0, 1.5) {prefix};
			\node[every text node part/.style={align=center}] (ut) at (0, 2) {uniquely\\decodable};
			\node (at) at (0, 2.5) {all codes};
			\node[every text node part/.style={align=center}] (ot) at (0, 1) {optimal\\prefix};

			\begin{scope}[shift={(1.4,0)}]
				\node [draw, circle, align=center, inner sep=1pt] (hl) at (-3.7, 1.25) {};
				\draw (hl.south east) -- (ht.west);
				\node[draw, ellipse, minimum width= 40pt, minimum height=20pt, align=center] (tl) at (-4.25, 1.25) {};
				\draw (tl.south) -- (tt.west);
				\node[draw, ellipse, minimum width= 11pt, minimum height=7pt, align=center] (ol) at (-3.6, 1.25) {};
				\draw (ol.east) -- (ot.west);
				\node [draw, ellipse, minimum width= 70pt, minimum height=35pt, align=center] (pl) at (-4, 1.25) {};
				\draw (pl.north east) -- (pt.west);
				\node [draw, ellipse, minimum width= 85pt, minimum height=45pt, align=center] (ul) at (-4, 1.25) {};
				\draw (ul.north east) -- (ut.west);
				\node [draw, rectangle, minimum width= 100pt, minimum height=55pt, align=center] (al) at (-4, 1.25) {};
				\draw (al.north east) -- (at.west);
				\node[above] at (al.north) {below tree line};
			\end{scope}

			\begin{scope}[shift={(-1.4,0)}]
				\node [draw, circle, align=center, inner sep=1pt] (hr) at (3.7, 1.25) {};
				\node[draw, ellipse, minimum width= 40pt, minimum height=20pt, align=center] (tr) at (4.125, 1.25) {};
				\node[draw, circle, align=center, inner sep=1pt] (or) at (3.125, 1.25) {};
				\node [draw, ellipse, minimum width= 70pt, minimum height=35pt, align=center] (pr) at (4, 1.25) {};
				\node [draw, ellipse, minimum width= 85pt, minimum height=45pt, align=center] (ur) at (4, 1.25) {};
				\node [draw, rectangle, minimum width= 100pt, minimum height=55pt, align=center] (ar) at (4, 1.25) {};
				\node [draw, ellipse, minimum width= 10pt, minimum height=5pt, align=center] (e) at (3.7, 1.25) {};
				\node[above] at (ar.north) {above tree line};
				\node (et) at (4.2, 0) {optimal tree-decodable};
			\end{scope}

			\draw (hr.south west) -- (ht.east);
			\draw (tr.south west) -- (tt.east);
			\draw (or.west) -- (ot.east);
			\draw (pr.north west) -- (pt.east);
			\draw (ur.north west) -- (ut.east);
			\draw (ar.north west) -- (at.east);
			\draw (e.south) -- (et.north);
		\end{tikzpicture}
	\vskip -.5em
	\caption{The relations between certain classes of source codes.}
	\label{fig:source}
	\ifnum\paperversion=1
	\vskip -2.3em
	\fi
\end{figure}
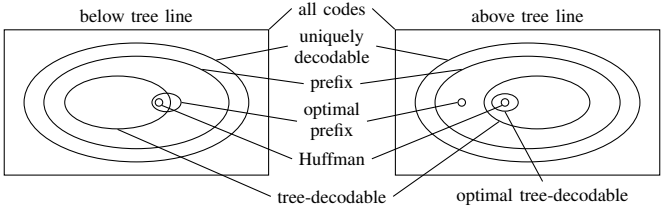

\ifnum\paperversion=2
As the alphabets of the channels can be different, we need to use a unified unit to measure the amount of information.
In this work, we use the unit ``nat'' (natural unit of information), i.e., we use $\ln$, the natural logarithm, in the evaluation of entropy.
All results in this work are valid if we use another base for the logarithm.
\fi

\begin{definition}[Lengths]
	The \emph{length tuple} of a $Q_n$-ary word $\mathbf{c}$ is %an $n$-tuple,
	denoted by $\len(\mathbf{c}) \coloneqq \langle \ell_0, \ell_1, \ldots, \ell_{n-1} \rangle = \langle \ell_i \rangle_{i \in \ZZ_n}$, where $\ell_i$ is the number of symbols in $\mathbf{c}(i)$.
	The \emph{$Q_n$-descriptive length} of a length tuple $\langle \ell_i \rangle_{i \in \ZZ_n}$ is defined as $|\langle \ell_i \rangle_{i \in \ZZ_n}|_{Q_n} \coloneqq \sum_{i \in \ZZ_n} \ell_i \ln q_i$. %\footnote{%
%As the alphabets of the channels can be different, we need to use a unified unit to measure the amount of information.
%In this work, we use the unit ``nat'' (natural unit of information), i.e., we use $\ln$, the natural logarithm, in the evaluation of entropy.
%All results in this work are valid if we use another base for the logarithm.
%}
	The \emph{descriptive length} of a $Q_n$-ary word $\mathbf{c}$ is defined as $|\mathbf{c}| \coloneqq |\len(\mathbf{c})|_{Q_n}$.
\end{definition}

The descriptive length of a word is the number of nats we need to represent the word.
The multichannel entropy bound \cite{yao10,packing} below states that the expected descriptive codeword length is no less than the entropy of the source, which is consistent with the single-channel one. %entropy bound.

If the finite sequences of codewords of any two distinct finite sequences of source symbols are different, then the source code is a \emph{uniquely decodable code}.
\ifnum\paperversion=2
Any codeword of a uniquely decodable code can be decoded without referring to the symbols of any future codewords if the code is tree-decodable \cite{yao10,packing}.
\fi

\noindent
\textbf{Entropy Bound:} For any uniquely decodable code %\footnote{%
%	If the finite sequences of codewords of any two distinct finite sequences of source symbols are different, then the source code is a \emph{uniquely decodable code}.
%	Any codeword of a uniquely decodable code can be decoded without referring to the symbols of any future codewords if the code is tree-decodable \cite{yao10,packing}.
%}
$\{\mathbf{c}_j\}_{j \in \ZZ_m}$ for a source random variable with probability $\{p_j\}_{j \in \ZZ_m}$, %and entropy $-\sum_{i \in \ZZ_m} p_i \ln p_i$, %the expected descriptive codeword length is no smaller than the entropy of $Z$, i.e.,
%\begin{equation*}
	$\sum_{j \in \ZZ_m} p_j |\mathbf{c}_j| \ge -\sum_{j \in \ZZ_m} p_j \ln p_j$.
%\end{equation*}
The equality holds if and only if $|\mathbf{c}_j| = -\ln p_j$, $\forall j \in \ZZ_m$.

\ifnum\paperversion=2
The value $|\mathbf{c}_j| + \ln p_j$ is the \emph{local redundancy} of the codeword $\mathbf{c}_j$ and the sum $\sum_{j \in \ZZ_m} p_j (|\mathbf{c}_j| + \ln p_j)$ is the \emph{redundancy} of the source code.
An \emph{optimal} code of a certain class of codes (e.g., tree-decodable codes, prefix codes, etc.) on a multiset $P$ of probabilities is a (symbol-by-symbol) code of that class having the lowest redundancy.
\else
An \emph{optimal} code of a certain class of codes (e.g., tree-decodable codes, prefix codes, etc.) on a multiset $P$ of probabilities is a (symbol-by-symbol) code of that class having the lowest redundancy, i.e., $\min \sum_{j \in \ZZ_m} p_j (|\mathbf{c}_j| + \ln p_j)$.
\fi

\begin{definition}[Tree Line]
	A tuple $Q$ of alphabet sizes is said to be \emph{above tree line} on a multiset of probabilities $P$ if there exists an optimal $Q$-ary tree-decodable code $C$ on $P$ such that $C$ is not an optimal $Q$-ary prefix code on $P$. If $Q$ is above tree line on some $P$, we say that $Q$ is above tree line, otherwise it is \emph{below tree line}.
\end{definition}

%We say $M$ is tree decodable if and only if $C$ is tree decodable.

As a brief summary, the relations between the classes of source codes we have mentioned are illustrated in \cref{fig:source}.

%\begin{definition}[$H$-Achieving]
%	A prefix code is \emph{entropy-achieving}, or in short, \emph{$H$-achieving}, if it achieves the entropy bound.
%\end{definition}

\subsection{Rectangle Packing Graph} \label{sec:rpg}

The \emph{rectangle packing graph (RPG)} \cite{packing,packing2} is a graphical tool to visualize the geometric nature of prefix codes.
%We first describe the general formulation of RPG.
%Then, we will give two examples to help the readers understand the formulation.
%
Consider a $Q_n$-ary source code $C$.
For each $\mathbf{c} \in C$, let $\len(\mathbf{c}) = \langle \ell_i^\mathbf{c} \rangle_{i \in \ZZ_n}$.
Let $\ellmax_i \coloneqq \max_{\mathbf{c} \in C} \ell_i^\mathbf{c}$.
The size of a $w_0 \times w_1 \times \ldots \times w_{n-1}$ hyper-rectangle is denoted by $\langle w_i \rangle_{i \in \ZZ_n}$.
%Denote by $\langle w_i \rangle_{i \in \ZZ_n}$ the size of a $w_0 \times w_1 \times \ldots \times w_{n-1}$ hyper-rectangle.

To draw an RPG for $C$, we first draw a \emph{container} $R$, which is a hyper-rectangle of size $\langle q_i^{\ellmax_i} \rangle_{i \in \ZZ_n}$.
For the $i$-th dimension, %the edge is a line comprises of integers $\{0, 1, 2, \ldots, q_i^{\ellmax_i}\}$ where each unit length corresponds to an integer in $\{0, 1, 2, \ldots, q_i^{\ellmax_i}-1\}$.\todo{rlai:This sentence makes no sense to me. What is a number line?}
the edge is an interval $[0, q_i^{\ellmax_i})$.
We write each integer $s$ in the above interval
%We write each of these integers $s$
in its unique $q_i$-ary numeral representation as a string of length $\lceil \log_{q_i}(s+1) \rceil$, then left-pad $0$'s to the string until the length becomes $\ellmax_i$.
That is, each unit hyper-cube in the RPG, which is called a \emph{cell}, corresponds to a word in $\prod_{i \in \ZZ_n} \ZZ_{q_i}^{\ellmax_i}$.

Next, each codeword $\mathbf{c}$ corresponds to a hyper-rectangle of size $\langle q_i^{\ellmax_i - \ell_i^\mathbf{c}} \rangle_{i \in \ZZ_n}$, which is called a \emph{block}.
The symbols of $\mathbf{c}$ specify the location where the block is put into the container.
Concretely, the block occupies exactly those cells whose common prefix of length $\len(\mathbf{c})$ is given by the codeword $\mathbf{c}$.

\begin{figure}
	%\negthickspace\negthickspace\negthickspace
	\begin{minipage}[t]{.5\textwidth}
	\centering
	\begin{minipage}[t]{.5\textwidth}
	\begin{figure}[H]
	\centering
	\scriptsize
	\begin{tikzpicture}
		\matrix (table) [matrix of math nodes,nodes={draw,minimum height=3ex,text width=3.5ex,align=center}] {
			000 & 001 & 010 & 011 & 100 & 101 & 110 & 111 \\
			};
			\begin{scope}[on background layer]
				\node[fill=red!30,inner xsep=0mm,inner ysep=0mm, fit=(table-1-1) (table-1-4) (table-1-4) (table-1-1)] {};
				\node[fill=blue!30,inner xsep=0mm,inner ysep=0mm, fit=(table-1-5) (table-1-6) (table-1-6) (table-1-5)] {};
				\node[fill=green!30,inner xsep=0mm,inner ysep=0mm, fit=(table-1-7) (table-1-7) (table-1-7) (table-1-7)] {};
			\end{scope}
	\end{tikzpicture}
	\vskip -.7em
	\caption{The RPG %of the binary codebook %$\{0, 10, 110\}$
		in \cref{eg:1drpg}.}
	\label{fig:1dtable}
	%\vskip -2.3em
	\end{figure}
	\end{minipage}~~
	\begin{minipage}[t]{.5\textwidth}
	\vskip -1em
	\begin{figure}[H]
	\centering
	\scriptsize
	\begin{tikzpicture}
		\matrix (table) [matrix of math nodes,nodes={draw,minimum height=3ex,text width=7ex,align=center}] {
			\substack{00\\1} & \substack{01\\1} & \substack{10\\1} & \substack{11\\1} \\
			\substack{00\\0} & \substack{01\\0} & \substack{10\\0} & \substack{11\\0} \\
			};
			\begin{scope}[on background layer]
				\node[fill=blue,opacity=.3,inner xsep=0mm,inner ysep=0mm, fit=(table-2-3) (table-2-4) (table-2-4) (table-2-3)] {};
				\node[fill=red,opacity=.3,inner xsep=0mm,inner ysep=0mm, fit=(table-1-1) (table-1-2) (table-2-2) (table-2-1)] {};
				\node[fill=green,opacity=.3,inner xsep=0mm,inner ysep=0mm, fit=(table-1-4) (table-1-4) (table-1-4) (table-1-4)] {};
			\end{scope}
	\end{tikzpicture}
	\vskip -.7em
	\caption{The RPG %of the $(2,2)$-ary codebook %$\{(0, \epsilon), (1, 0), (11, 1)\}$
		in \cref{eg:2drpg}.}
	\label{fig:2dtable}
	\end{figure}
	\end{minipage}
	\end{minipage}
	\ifnum\paperversion=1
	\vskip -2em
	\fi
\end{figure}

\begin{example} \label{eg:1drpg}
	The RPG of a binary %single-channel
	codebook $\{0, 10, 110\}$ is illustrated in \cref{fig:1dtable}.
	The container is of length $8$.
	%Technically the container is a $1$-D line, but we draw it in $2$-D for the ease of presentation.
	The blocks for $0$ (red), $10$ (blue) and $110$ (green) have length $4$, $2$ and $1$ respectively.
	%$2^{(3-1)} = 4$, $2^{(3-2)} = 2$ and $2^{(3-3)} = 1$ respectively.
\end{example}

%\begin{figure}[b]
%	\vskip -2em
%	\centering
%	\scriptsize
%	\begin{tikzpicture}
%		\matrix (table) [matrix of math nodes,nodes={draw,minimum height=3ex,text width=7ex,align=center}] {
%			\substack{00\\1} & \substack{01\\1} & \substack{10\\1} & \substack{11\\1} \\
%			\substack{00\\0} & \substack{01\\0} & \substack{10\\0} & \substack{11\\0} \\
%			};
%			\begin{scope}[on background layer]
%				\node[fill=blue,opacity=.3,inner xsep=0mm,inner ysep=0mm, fit=(table-2-3) (table-2-4) (table-2-4) (table-2-3)] {};
%				\node[fill=red,opacity=.3,inner xsep=0mm,inner ysep=0mm, fit=(table-1-1) (table-1-2) (table-2-2) (table-2-1)] {};
%				\node[fill=green,opacity=.3,inner xsep=0mm,inner ysep=0mm, fit=(table-1-4) (table-1-4) (table-1-4) (table-1-4)] {};
%			\end{scope}
%	\end{tikzpicture}
%	\caption{The RPG of the $(2,2)$-ary codebook $\{(0, \epsilon), (1, 0), (11, 1)\}$ in \cref{eg:2drpg}.}
%	\label{fig:2dtable}
%\end{figure}

\begin{example} \label{eg:2drpg}
	The RPG of a $(2,2)$-ary codebook $\{(0,\epsilon)$, $(1,0)$, $(11,1)\}$ is illustrated in \cref{fig:2dtable}, where the horizontal and vertical axes corresponds to the 0-th and 1-st channels respectively.
	The container is a $4 \times 2$ rectangle.
	%$\langle 2^2, 2^1 \rangle = \langle 4, 2 \rangle$ rectangle.
	The blocks for $(0, \epsilon)$ (red), $(1, 0)$ (blue) and $(11, 1)$ (green) are $2 \times 2$, $2 \times 1$ and $1 \times 1$ rectangles respectively.
	%$\langle 2^{(2-1)}, 2^{1-0} \rangle = \langle 2, 2 \rangle$, $\langle 2^{2-1}, 2^{1-1} \rangle = \langle 2, 1 \rangle$ and $\langle 2^{2-2}, 2^{1-1} \rangle = \langle 1, 1 \rangle$ rectangles respectively.
\end{example}

All the blocks $b_\mathbf{c}$ for $\mathbf{c} \in C$ do not overlap with each other if and only if $C$ is a prefix code \cite{packing}.
We can write $b_\mathbf{c} \cap b_{\mathbf{c}'} = \emptyset$ to indicate that these two blocks do not overlap, but we cannot write $\mathbf{c} \cap \mathbf{c}' = \emptyset$ as codewords are not sets.
So, we use the notation $\mathbf{c} \prefix \mathbf{c}'$,
which symbolizes that the corresponding blocks do not overlap.

Suppose $C$ is a prefix code.
Let $\vol(R) = \prod_{i \in \ZZ_n} q_i^{\ellmax_i}$ be the hyper-volume of the container.
The hyper-volume of the block $b_\mathbf{c}$ is $\vol(\mathbf{c}) = \vol(R) \exp(-|\mathbf{c}|)$.
%Let $\vol(\mathbf{c})$ be the hyper-volume of the block $b_\mathbf{c}$, which equals $\prod_{i \in \ZZ_n} q_i^{\ellmax_i - \ell_i^\mathbf{c}} = \vol(R) \exp(-|\mathbf{c}|)$.
As the blocks do not overlap, the sum of volumes of the blocks must be no larger than $\vol(R)$.
%the sum of their volumes must be less than the volume of the container.
This coincides with the multichannel Kraft inequality \cite{yao10,packing}. %after normalization.
%Normalizing $\vol(R)$, we get the multichannel Kraft inequality \cite{yao10,packing}.
%Note that the Kraft inequality actually applies to any uniquely decodable code.

\noindent
\textbf{Kraft Inequality:} For any uniquely decodable code $\{\mathbf{c}_j\}_{j \in \ZZ_m}$, the descriptive lengths of the codewords satisfy
%\begin{equation*}
	$\sum_{j \in \ZZ_m} \exp(-|\mathbf{c}_j|) \le 1$.
%\end{equation*}

%However, unlike its single-channel counterpart, satisfying the Kraft inequality does not imply the existence of prefix codes as only volumes but not the geometry of the blocks are considered.
%A minimal example is a $(2,2)$-ary non-prefix code $\{(0,\epsilon), (\epsilon,0)\}$, where its RPG is a $2 \times 2$ container with a $1 \times 2$ and a $2 \times 1$ block.

\section{Non-Tree-Decodable Prefix Codes}

\subsection{Guillotine-Cuts and Tree Decodability}

In the view of RPG, a decoding tree is a variant of a k-d tree constructed as follows.
The root node corresponds to the container and is assigned a class $i \in \ZZ_n$.
Each internal or leaf node, corresponding to a subspace of the container, is constructed recursively using the following procedure:
For each node of class $i$, we guillotine-cut the space corresponding to the node by $(q_i-1)$ hyper-planes perpendicular to the $i$-th dimension into $q_i$ subspaces of equal size, and assign a node to each subspace.
Upon completion, each leaf corresponds to a block, i.e., a codeword.
Summarizing, the decoding tree is a k-d tree formed by performing equal-space partitioning where the number of children and the orientation depend on the class of the node.

\begin{example}
	We take the RPG in \cref{eg:2drpg} (\cref{fig:2dtable}) as an example.
	The decoding tree is illustrated in \cref{fig:2dtree}.
	The number in each non-leaf node is the class of the node.
	Each color of the branches corresponds to the guillotine-cuts with the same color in the RPG below the tree.
	The orientation of each guillotine-cut depends on the class of the node.
\end{example}

\begin{theorem} \label{thm:guillotine}
	A prefix code is tree decodable if and only if all the blocks representing the codewords in the rectangle packing graph can be obtained by guillotine-cuts.
\end{theorem}

%\begin{IEEEproof}
%	If a prefix code is not tree decodable, then it means that there is a subspace $S$ such that $S$ is guillotine-cut from the container, or $S$ is the container itself, such that $S$ is occupied by more than one block but no guillotine-cut is possible without cutting through a block.
%	That is, there is no way to obtain any block in $S$ by guillotine-cuts.
%
%	Conversely, suppose some blocks cannot be obtained by guillotine-cuts.
%	We consider a subset $T$ of these blocks having the same prefix where $|T| > 1$ such that the smallest space $S$ containing the blocks in $T$ is not possible to be separated by guillotine-cuts without cutting through a block.
%	Suppose the prefix code is tree decodable, then there is a non-leaf node corresponds to the space $S$.
%	However, there is no way to further guillotine-cut $S$, so we cannot find a class for this node, which contradicts that the code is tree decodable.
%\end{IEEEproof}

%\subsection{$4$-D and $5$-D RPGs}
\subsection{Interweave Structures}

\begin{figure}
	%\vskip -1.2em
	\centering
	\tiny
	\begin{subfigure}{.15\textwidth}
		\centering
		\begin{tikzpicture}[scale=.4]
			\useasboundingbox (-3,-1.5) rectangle (3,1.5);
			\draw (0,0) ellipse (3 and 1.5);
			\begin{scope}
				\clip (0,-1.8) ellipse (3 and 2.5);
				\draw (0,2.2) ellipse (3 and 2.5);
			\end{scope}
			\begin{scope}
				\clip (0,2.2) ellipse (3 and 2.5);
				\draw (0,-2.2) ellipse (3 and 2.5);
			\end{scope}
			\begin{scope}
				\clip (1.5,.1) rectangle (4,2);
				\draw (0,0) ellipse (3.2 and 1.7);
			\end{scope}
			\draw[-stealth] (3.2,.1) -- ++(0,-.1);
			\node at (3.4,1.0) [rotate=-45,every text node part/.style={align=center}] {2-nd and 3-rd\\channels};
			\draw[-stealth] ($(-2.25,0)+(-.7071,-.7071)$) arc (225:30:1);
			\node at (-2.3,1.7) [every text node part/.style={align=center}] {0-th and 1-st\\channels};
		\end{tikzpicture}
		\vskip -.6em
		\caption{The bicycle tube formed after merging the 0-th and the 1-st channels, and merging the 2-nd and the 3-rd channels.}
		\label{fig:3d4a}
	\end{subfigure}~
	\begin{subfigure}{.35\textwidth}
		\centering
		\begin{tikzpicture}[scale=.3]
			\foreach \r in {1, 2, 3}
			\draw (0,0) circle (\r);
			\draw[-stealth] (1,0) -- node[midway,sloped,below] {4-th channel} (3,0);
			\draw[-stealth] (-1,0) -- node[midway,sloped,below] {4-th channel} (-3,0);
			\node at (1.5,0) [above] {$0$};
			\node at (2.5,0) [above] {$1$};
			\node at (-1.5,0) [above] {$0$};
			\node at (-2.5,0) [above] {$1$};
			\draw[ultra thick,green,opacity=.3] (45:1) -- (45:3);
			\draw[ultra thick,red,opacity=.3] (-135:1) -- (-135:3);
			\draw[ultra thick,blue,opacity=.3] (135:1) -- (135:3);
			\draw[ultra thick,black,opacity=.3] (-45:1) -- (-45:3);
			\node at (145:3.1) [left] {$00$};
			\node at (125:3.1) [above] {$01$};
			\node at (-35:3.1) [right] {$11$};
			\node at (-55:3.1) [below] {$10$};
			\draw (225:3.7) arc (225:180:3.7) node[above,rotate=90,every text node part/.style={align=center}] {0-th and 1-st\\channels};
			\draw[-stealth] (180:3.7) arc (180:160:3.7);

			\begin{scope}[shift={(8,0)}]
				\foreach \r in {1, 2, 3}
				\draw (0,0) circle (\r);
				\draw[-stealth] (1,0) -- node[midway,sloped,below] {4-th channel} (3,0);
				\draw[-stealth] (-1,0) -- node[midway,sloped,below] {4-th channel} (-3,0);
				\node at (1.5,0) [above] {$0$};
				\node at (2.5,0) [above] {$1$};
				\node at (-1.5,0) [above] {$0$};
				\node at (-2.5,0) [above] {$1$};
				\draw[ultra thick,blue,opacity=.3] (45:1) -- (45:3);
				\draw[ultra thick,black,opacity=.3] (-135:1) -- (-135:3);
				\draw[ultra thick,green,opacity=.3] (135:1) -- (135:3);
				\draw[ultra thick,red,opacity=.3] (-45:1) -- (-45:3);
				\node at (35:3.1) [right] {$00$};
				\node at (55:3.1) [above] {$01$};
				\node at (-145:3.1) [left] {$11$};
				\node at (-125:3.1) [below] {$10$};
				\draw (-45:3.7) arc (-45:0:3.7) node[above,rotate=-90,every text node part/.style={align=center}] {0-th and 1-st\\channels};
				\draw[-stealth] (0:3.7) arc (0:20:3.7);
			\end{scope}
		\end{tikzpicture}
		\vskip -.6em
		\caption{One of the cross sections of the bicycle tube.}
		\label{fig:3d4b}
	\end{subfigure}\\
	\begin{subfigure}{.23\textwidth}
		\centering
		\tdplotsetmaincoords{82}{40}
		\begin{tikzpicture}[scale=.5]
			\foreach \z in {0, 1, 2, 3, 4}{
				\begin{scope}[tdplot_main_coords,canvas is xy plane at z=\z]
					\foreach \r in {1, 2, 3}
					\draw (0,0) circle (\r);
				\end{scope}
			}
			\foreach \x in {1, 2, 3}{
				\begin{scope}[tdplot_main_coords]
					\draw (\x,0,0) -- ++(0,0,4);
					\draw (-\x,0,0) -- ++(0,0,4);
					\draw (0,\x,0) -- ++(0,0,4);
					\draw (0,-\x,0) -- ++(0,0,4);
				\end{scope}
			}
			\foreach \z in {0, 1, 2, 3, 4}{
				\begin{scope}[tdplot_main_coords]
					\draw (1,0,\z) -- ++(2,0,0);
					\draw (-1,0,\z) -- ++(-2,0,0);
					\draw (0,1,\z) -- ++(0,2,0);
					\draw (0,-1,\z) -- ++(0,-2,0);
				\end{scope}
			}
			\foreach \i in {1, 2, 3}{
				\pgfmathsetmacro{\xx}{\i*cos(40)}
				\pgfmathsetmacro{\yy}{\i*sin(40)}
				\draw[dashed,tdplot_main_coords] (\xx,\yy,0) -- ++(0,0,4);
				\draw[dashed,tdplot_main_coords] (-\xx,-\yy,0) -- ++(0,0,4);
			}
			\begin{scope}[tdplot_main_coords]
				\pgfmathsetmacro{\xx}{3*cos(40)}
				\pgfmathsetmacro{\yy}{3*sin(40)}
				\coordinate (z0) at (\xx,\yy,0);
				\coordinate (z1) at (\xx,\yy,4);
			\end{scope}

			\fill[tdplot_main_coords,blue,opacity=.3] (-3,0,0) -- ++(2,0,0) -- ++(0,0,4) -- ++(-2,0,0) -- cycle;
			\fill[tdplot_main_coords,green,opacity=.3] (0,3,0) -- ++(0,-2,0) -- ++(0,0,4) -- ++(0,2,0) -- cycle;
			\fill[tdplot_main_coords,red,opacity=.3] (0,-3,0) -- ++(0,2,0) -- ++(0,0,4) -- ++(0,-2,0) -- cycle;
			\fill[tdplot_main_coords,black,opacity=.3] (3,0,0) -- ++(-2,0,0) -- ++(0,0,4) -- ++(2,0,0) -- cycle;

			\draw[-stealth] ($(z0)+(.2,0)$) -- node[midway,sloped,below] {2-nd and 3-rd channels} ($(z1)+(.2,0)$);
		\end{tikzpicture}
		\vskip -.6em
		\caption{The cylinder after cutting at one side of the cross section shown in \cref{fig:3d4b}.}
		\label{fig:3d4c}
	\end{subfigure}~~~~
	\begin{subfigure}{.23\textwidth}
		\centering
		\tdplotsetmaincoords{70}{30}
		\begin{tikzpicture}[scale=.5]
			\begin{scope}[tdplot_main_coords]
				\foreach \x in {0,1,2,3,4}
				\foreach \y in {0,1,2,3,4}
				\foreach \z in {0,1,2}
				{
					\draw (\x,0,\z) -- (\x,4,\z);
					\draw (0,\y,\z) -- (4,\y,\z);
					\draw (\x,\y,0) -- (\x,\y,2);
				}

				\fill[red,opacity=.3] (0,0,0) -- ++(0,4,0) -- ++(0,0,2) -- ++(0,-4,0) -- cycle;
				\fill[blue,opacity=.3] (1,0,0) -- ++(0,4,0) -- ++(0,0,2) -- ++(0,-4,0) -- cycle;
				\fill[green,opacity=.3] (2,0,0) -- ++(0,4,0) -- ++(0,0,2) -- ++(0,-4,0) -- cycle;
				\fill[black,opacity=.3] (3,0,0) -- ++(0,4,0) -- ++(0,0,2) -- ++(0,-4,0) -- cycle;
				\fill[red,opacity=.3] (4,0,0) -- ++(0,4,0) -- ++(0,0,2) -- ++(0,-4,0) -- cycle;

				\coordinate (x0) at (0,0,0);
				\coordinate (y0) at (4,0,0);
				\coordinate (z0) at (4,4,0);
				\coordinate (z1) at (4,4,2);

				\coordinate (x1) at (.5,4,2);
				\coordinate (x2) at (1.5,4,2);
				\coordinate (x3) at (2.5,4,2);
				\coordinate (x4) at (3.5,4,2);

				\coordinate (y1) at (0,.5,2);
				\coordinate (y2) at (0,1.5,2);
				\coordinate (y3) at (0,2.5,2);
				\coordinate (y4) at (0,3.5,2);
			\end{scope}
			\draw[-stealth] ($(x0)+(0,-.1)$) -- node[midway,sloped,below] {0-th and 1-st channels} ($(y0)+(0,-.1)$);
			\draw[-stealth] ($(y0)+(0,-.1)$) -- node[midway,sloped,below] {2-nd and 3-rd channels} ($(z0)+(.1,0)$);
			\draw[-stealth] ($(z0)+(.1,0)$) -- node[midway,sloped,below] {4-th channel} ($(z1)+(.1,0)$);
			\node at ($(x1)+(0,.2)$) {$00$};
			\node at ($(x2)+(0,.2)$) {$01$};
			\node at ($(x3)+(0,.2)$) {$11$};
			\node at ($(x4)+(0,.2)$) {$10$};
			\node at ($(y1)+(-.2,.1)$) {$00$};
			\node at ($(y2)+(-.2,.1)$) {$01$};
			\node at ($(y3)+(-.2,.1)$) {$11$};
			\node at ($(y4)+(-.2,.1)$) {$10$};
		\end{tikzpicture}
		\vskip -.6em
		\caption{The rectangle formed after further cutting the cylinder at the red plane.}
		\label{fig:3d4d}
	\end{subfigure}
	\vskip -.8em
	\caption{The representation of a $5$-D RPG by a $3$-D rectangle.}
	\label{fig:5drpg}
	\ifnum\paperversion=1
	\vskip -3.5em
	\fi
\end{figure}
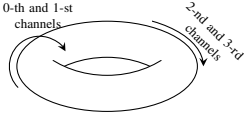
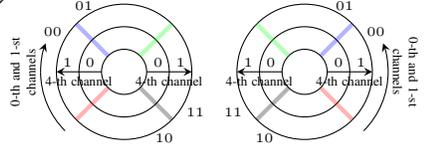
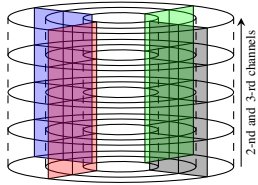
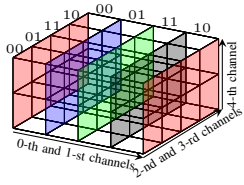

We need to use a more-than-$3$-D RPG in the following discussion. %, which have not been studied in previous works.
To illustrate a $4$-D hyper-rectangle of size $\langle 2 \rangle_{i \in \ZZ_4}$, we concatenate the 0-th and the 1-st channels into one dimension and use a Gray code to let consecutive symbols of the same channel adjacent to each other.
Then, we can transform the RPG into a cuboid as shown in \cref{fig:4drpg}.\footnote{%
	The RPG is a cylinder with a hole in the middle as shown in \cref{fig:3d2}.
	When we perform a guillotine-cut on the 0-th channel, we need to cut through both sides of the cylinder, i.e., cut both the red planes in the figure.
	Similarly, we need to cut both blue planes to guillotine-cut the 1-st channel.
	For a simpler illustration, we can cut one of the red plane and bend the structure into a rectangle as shown in \cref{fig:3d2a}.
	This way, each guillotine-cut on the 0-th or the 1-st channel in the $4$-D space becomes two guillotine-cuts in this rectangle.
}
	%Now, we illustrate in \cref{fig:3d3} the RPG of the above codewords in a $3$-D rectangle.
	%We can see that although we only have three codewords, they can already form an interweave structure for four channels.
%\end{example}
%
For the $5$-D hyper-rectangle of size $\langle 2 \rangle_{i \in \ZZ_5}$, besides merging the first two channels, we merge the 2-nd and the 3-rd channels by a Gray code.
Similarly, we can transform the RPG into a cuboid as shown in \cref{fig:5drpg}.\footnote{%
	The RPG is a bicycle tube: a torus with thickness and with a hollow tube inside.
	\cref{fig:3d4a,fig:3d4b} illustrate the bicycle tube and one of its cross sections respectively.
	By cutting \cref{fig:3d4a} at one side of the cross section shown in \cref{fig:3d4b}, we bend the tube into a cylinder structure as in the $4$-D RPG shown in \cref{fig:3d4c}.
	By further cutting one side (the red plane) of the cylinder, we can bend the structure into a rectangle shown in \cref{fig:3d4d}.
}
%The main message is that, for a guillotine-cut on a cuboid representation of a higher dimensional RPG, we may need to cut multiple times on the $3$-D rectangle, e.g., the red planes in \cref{fig:3d2a} corresponds to a single cut.
%Also, 
Note that 
a block can be sheared into subblocks. %, e.g., a block for the codeword $(\epsilon, 0, 0, 0, 0)$ passes through (wrap around) the red planes in \cref{fig:3d4d}. %\footnote{%
%	When the dimensions merged by Gray codes are of different lengths, we can still draw a $3$-D representation by merging the channels, but a block would be further sheared into more subblocks as we do not have an easy way to preserve the continuity of a higher-dimensional object.
%}

%\subsection{Interweave Structures}

We consider a space containing more than one block such that no guillotine-cut is possible without cutting through a block.
For each dimension, there must be a block of length matching the one of the space at this dimension, or otherwise we can further guillotine-cut the space due to the constraints on the location and the size of the blocks.
By removing the common prefix of the blocks, we can treat the space as the container itself and thus the blocks form a prefix code.
If there is a dimension such that the length of every block in this dimension matches the one of the space, we can safely remove this whole dimension.
This dimension corresponds to a \emph{dummy channel}, which is unused by any codeword. %\footnote{%
	%After removing all dummy channels, we must have at least $3$ channels remaining as a non-tree-decodable prefix code.
%}

In the following discussion, we regard the resulting space after performing the actions described above as the container, i.e., the common prefix is removed from all the codewords and then the dummy channels are removed.
Let $T$ be the set of all codewords in this space and $t$ be the number of dimensions of this space.
Without loss of generality, assume these $t$ channels are the first $t$ channels.

\begin{definition}
	An \emph{$\epsilon$-locating function} for an $m \times n$ coding matrix $M$ is a function $\mathcal{E}_M \colon \ZZ_n \to 2^{\ZZ_m}$ defined as $\mathcal{E}_M(i) = \{j \colon M_{j,i} = \epsilon\}$.
\end{definition}

\begin{theorem} \label{thm:E}
	Let $M$ be an $n$-column codeword matrix.
	If $\mathcal{E}_M(i) \neq \emptyset$ for all $i \in \ZZ_n$, then $M$ is not tree decodable.
\end{theorem}

%\begin{IEEEproof}
%	Suppose $M$ is tree decodable.
%	The root node of the decoding tree must belong to one of the classes in $\ZZ_n$.
%	However, for each possible class $k$, there exists at least one codeword that is not a descendant of the root node.
%	This contradicts that $M$ is tree decodable.
%\end{IEEEproof}

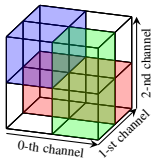
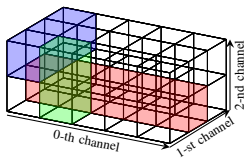
\begin{figure}[b]
	\ifnum\paperversion=1
	\vskip -2em
	\fi
	\centering
	\begin{subfigure}{.2\textwidth}
		\centering
		\tiny
		\tdplotsetmaincoords{70}{20}
		\begin{tikzpicture}[scale=.65]
			\begin{scope}[tdplot_main_coords]
				\foreach \x in {0,1,2}
				\foreach \y in {0,1,2}
				\foreach \z in {0,1,2}
				{
					\draw (\x,0,\z) -- (\x,2,\z);
					\draw (0,\y,\z) -- (2,\y,\z);
					\draw (\x,\y,0) -- (\x,\y,2);
				}
				\fill[red,opacity=.3] (0,1,0) -- ++(2,0,0) -- ++(0,1,0) -- ++(0,0,1) -- ++(-2,0,0) -- ++(0,-1,0) -- cycle;
				\fill[blue,opacity=.3] (0,0,1) -- ++(1,0,0) -- ++(0,2,0) -- ++(0,0,1) -- ++(-1,0,0) -- ++(0,-2,0) --cycle;
				\fill[green,opacity=.3] (1,0,0) -- ++(1,0,0) -- ++(0,1,0) -- ++(0,0,2) -- ++(-1,0,0) -- ++(0,-1,0) -- cycle;
				\coordinate (x0) at (0,0,0);
				\coordinate (y0) at (2,0,0);
				\coordinate (z0) at (2,2,0);
				\coordinate (z1) at (2,2,2);
			\end{scope}
			\draw[-stealth] ($(x0)+(0,-.1)$) -- node[midway,sloped,below] {0-th channel} ($(y0)+(0,-.1)$);
			\draw[-stealth] ($(y0)+(0,-.1)$) -- node[midway,sloped,below] {1-st channel} ($(z0)+(.1,0)$);
			\draw[-stealth] ($(z0)+(.1,0)$) -- node[midway,sloped,below] {2-nd channel} ($(z1)+(.1,0)$);
		\end{tikzpicture}
		\vskip -.8em
		\caption{$Q = (2, 2, 2)$.}
		\label{fig:222core}
	\end{subfigure}~
	\begin{subfigure}{.3\textwidth}
		\centering
		\tiny
		\tdplotsetmaincoords{70}{30}
		\begin{tikzpicture}[scale=.5]
			\begin{scope}[tdplot_main_coords]
				\foreach \x in {0,1,2,3,4,5}
				\foreach \y in {0,1,2,3}
				\foreach \z in {0,1,2}
				{
					\draw (\x,0,\z) -- (\x,3,\z);
					\draw (0,\y,\z) -- (5,\y,\z);
					\draw (\x,\y,0) -- (\x,\y,2);
				}
				%\fill[red!40!black,opacity=.3] (0,1,0) -- ++(5,0,0) -- ++(0,1,0) -- ++(-5,0,0) -- cycle;
				%\fill[red,opacity=.3] (0,2,0) -- ++(5,0,0) -- ++(0,0,1) -- ++(-5,0,0) -- cycle;
				%\fill[red,opacity=.3] (0,1,0) -- ++(0,1,0) -- ++(0,0,1) -- ++(0,-1,0) -- cycle;
				%\fill[red!80!black,opacity=.3] (0,1,1) -- ++(5,0,0) -- ++(0,1,0) -- ++(-5,0,0) -- cycle;
				%\fill[red!80!white,opacity=.3] (0,1,0) -- ++(5,0,0) -- ++(0,0,1) -- ++(-5,0,0) -- cycle;
				%\fill[red!50!white,opacity=.3] (5,1,0) -- ++(0,1,0) -- ++(0,0,1) -- ++(0,-1,0) -- cycle;
				\fill[red,opacity=.3] (0,1,0) -- ++(5,0,0) -- ++(0,1,0) -- ++(0,0,1) -- ++(-5,0,0) -- ++(0,-1,0) -- cycle;
				\fill[blue,opacity=.3] (0,0,1) -- ++(1,0,0) -- ++(0,3,0) -- ++(0,0,1) -- ++(-1,0,0) -- ++(0,-3,0) --cycle;
				\fill[green,opacity=.3] (1,0,0) -- ++(1,0,0) -- ++(0,1,0) -- ++(0,0,2) -- ++(-1,0,0) -- ++(0,-1,0) -- cycle;
				\coordinate (x0) at (0,0,0);
				\coordinate (y0) at (5,0,0);
				\coordinate (z0) at (5,3,0);
				\coordinate (z1) at (5,3,2);
			\end{scope}
			\draw[-stealth] ($(x0)+(0,-.1)$) -- node[midway,sloped,below] {0-th channel} ($(y0)+(0,-.1)$);
			\draw[-stealth] ($(y0)+(0,-.1)$) -- node[midway,sloped,below] {1-st channel} ($(z0)+(.1,0)$);
			\draw[-stealth] ($(z0)+(.1,0)$) -- node[midway,sloped,below] {2-nd channel} ($(z1)+(.1,0)$);
		\end{tikzpicture}
		\vskip -.8em
		\caption{$Q = (5, 3, 2)$.}
		\label{fig:532core}
	\end{subfigure}
	\vskip -.4em
	\caption{The RPG of the $Q$-ary codebook $\{(\epsilon,1,0), (0,\epsilon,1), (1,0,\epsilon)\}$ in \cref{eg:trim}.}
	\label{fig:3d1}
	%\vskip -1.8em
\end{figure}

\begin{example} \label{eg:trim}
	Suppose the space which cannot be guillotine-cut contains the codewords $(1, 01, 10, 0)$, $(10, 0, 11, 0)$ and $(11, 00, 1, 0)$.
	By removing the common prefix, the codewords become $(\epsilon, 1, 0, \epsilon)$, $(0, \epsilon, 1, \epsilon)$ and $(1, 0, \epsilon, \epsilon)$.
	%The last channel is \emph{dummy} as it is unused. %, i.e., every block has the same length in the last dimension which matches the one of the space.
	By removing the dummy channel, the codewords become $(\epsilon, 1, 0)$, $(0, \epsilon, 1)$ and $(1, 0, \epsilon)$.
By \cref{thm:E}, these codewords cannot form a decoding tree.
%Note that these codewords form the simplest structure in the sense that $|\mathcal{E}_M(j)| = 1$ for all $j$, i.e., each column in the codeword matrix consists of exactly one $\epsilon$.
This fact can be visualized in the RPG of this code illustrated in \cref{fig:3d1}.
%We can see that each block in this RPG has one and only one edge fitting the one of the container in the same dimension.
%Each block looks like a selvage at the edges of denim, so we call this code the \emph{selvage core} and we defer further discussion on this core to the next section.
The blocks interweave with each other so that it is impossible to separate these blocks using guillotine-cuts.
This is one of the interweave structures which makes prefix codes not tree decodable.
\end{example}

Note that it is not necessary to have $t$ interweaving blocks to form a $t$-channel non-tree-decodable prefix code.
Also, we can have more than one $\epsilon$ per column or per row in the codeword matrix.
To demonstrate these, we need to consider a higher dimensional hyper-rectangle.
%We give two more examples below.
Below is an example using a $5$-D RPG.

\begin{example} \label{eg:5d}
	Consider a $(2,2,2,2,2)$-ary codeword matrix
	\begin{equation*}
		\begin{matrix}
			\text{\tiny red}\\
			\text{\tiny blue}\\
			\text{\tiny green}\\
			\text{\tiny yellow}\\
			\text{\tiny gray}
		\end{matrix}
		\begin{pmatrix}
			\epsilon & \epsilon & 1 & 0 & 0\\
			0 & \epsilon & \epsilon & 1 & 0\\
			0 & 0 & \epsilon & \epsilon & 1\\
			1 & 0 & 0 & \epsilon & \epsilon\\
			\epsilon & 1 & 0 & 0 & \epsilon
		\end{pmatrix}.
	\end{equation*}
	We have two $\epsilon$'s per row and per column.
	%By using a similar technique as in \cref{eg:4d}, we merge the 0-th and 1-st channels together and merge the 2-nd and 3-rd channels together.
	The RPG is illustrated in \cref{fig:3d4}.
	%We can see that the interweave structure is more complicated than drawing three perpendicular blocks.
	We can see that if we remove the blue and the yellow blocks, the remaining three blocks can still make the $5$-D container not guillotine-cutable, i.e., the code is still not tree decodable.
	In this new codeword matrix, we still have two $\epsilon$'s in the first column, one $\epsilon$ in each other column, and each row has two $\epsilon$'s.
\end{example}

\section{Main Result}

In this section, we will show a sufficient condition for the channel alphabets $Q$ being above tree line.
%we prove that if $Q$ satisfies a property called ``$t$-separable'' for $t \geq 3$, then $Q$ is above tree line. Roughly, $Q$ is $t$-separable if it can be partitioned into $t$ chunks $\{Q_j\}_{j \in \ZZ_t}$ called a $t$-separation which satisfies certain properties.
%
At the core of our result is the construction of a $Q$-ary prefix code, called the \emph{$Q$-ary selvage code}, %\footnote{%
%	The RPG of a selvage code has a simple interweave structure:
%	Each block in the interweave core has exactly one edge fitting an edge of the container in the same dimension.
%	The RPGs in \cref{fig:3d1} fall in this category.
%	For each plane, this structure looks like a selvage at the edges of denim, so we name our construction``selvage''.
%},
which achieves the entropy bound on a special multiset of probabilities called the \emph{$Q$-ary selvage probability assembly (SPA)}.
%The $Q$-ary selvage code is designed in a way that no $Q$-ary prefix code with the same codeword lengths is tree-decodable.
%%The RPG of a selvage code has a simple interweave structure:
%%Each block in the interweave core has exactly one edge fitting an edge of the container in the same dimension.
%%The RPGs in \cref{fig:3d1} fall in this category.
%%For each plane, this structure looks like a selvage at the edges of denim, so we name our construction ``selvage''.
%Given the construction of $Q$-ary selvage codes, proving the sufficient condition boils down to proving a ``separable'' property of $Q$.

%that, if $Q$ is $t$-separable for some $t \geq 3$, then any optimal $Q$-ary prefix code on the $Q'$-ary SPA must have the same codeword lengths as those of the $Q'$-ary selvage code, where $Q'$ is the ``product alphabets'' $Q' = (\prod_{q \in Q_j} q)_{j \in \ZZ_t}$ induced by the $t$-separation of $Q$.

\subsection{Selvage Code and Selvage Probability Assembly (SPA)}

We now construct the $Q$-ary selvage code and then assign probabilities, i.e., the $Q$-ary SPA, to the codewords so that the code achieves the entropy (bound) on these probabilities.

To ensure that the $Q$-ary selvage code is not tree-decodable, our strategy is to construct its codeword matrix such that it consists of an $n \times n$ cyclic codeword submatrix as the interweave core, called the \emph{selvage core}, where the diagonal is filled with $\epsilon$, the off-diagonal above the main one is filled with $1$ cyclically, and the remaining entries are $0$.
Then, we assign probability $p_j$ to the $j$-th codeword in the core such that the local redundancy $|\mathbf{c}_j| + \ln p_j$ is zero.
Concretely, we set $p_j \coloneqq q_j \prod_{i \in \ZZ_n} q_i^{-1}$.

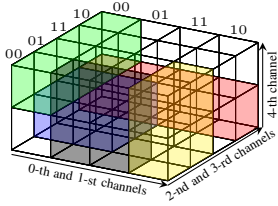
\begin{figure}
	\negthickspace
	\begin{minipage}[t]{.5\textwidth}
	\centering
	\begin{minipage}[t]{.5\textwidth}
	\vskip -1em
	\begin{figure}[H]
	\centering
	\tiny
	\tdplotsetmaincoords{70}{30}
	\begin{tikzpicture}[scale=.6]
		\begin{scope}[tdplot_main_coords]
			\foreach \x in {0,1,2,3,4}
			\foreach \y in {0,1,2,3,4}
			\foreach \z in {0,1,2}
			{
				\draw (\x,0,\z) -- (\x,4,\z);
				\draw (0,\y,\z) -- (4,\y,\z);
				\draw (\x,\y,0) -- (\x,\y,2);
			}

			\fill[red,opacity=.3] (0,3,0) -- ++(4,0,0) -- ++(0,1,0) -- ++(0,0,1) -- ++(-4,0,0) -- ++(0,-1,0) -- cycle;
			\fill[blue,opacity=.3] (0,1,0) -- ++(2,0,0) -- ++(0,2,0) -- ++(0,0,1) -- ++(-2,0,0) -- ++(0,-2,0) -- cycle;
			\fill[green,opacity=.3] (0,0,1) -- ++(1,0,0) -- ++(0,4,0) -- ++(0,0,1) -- ++(-1,0,0) -- ++(0,-4,0) -- cycle;
			\fill[black,opacity=.4] (1,0,0) -- ++(2,0,0) -- ++(0,1,0) -- ++(0,0,2) -- ++(-2,0,0) -- ++(0,-1,0) -- cycle;
			\fill[yellow,opacity=.3] (3,0,0) -- ++(1,0,0) -- ++(0,2,0) -- ++(0,0,2) -- ++(-1,0,0) -- ++(0,-2,0) -- cycle;

			\coordinate (x0) at (0,0,0);
			\coordinate (y0) at (4,0,0);
			\coordinate (z0) at (4,4,0);
			\coordinate (z1) at (4,4,2);

			\coordinate (x1) at (.5,4,2);
			\coordinate (x2) at (1.5,4,2);
			\coordinate (x3) at (2.5,4,2);
			\coordinate (x4) at (3.5,4,2);

			\coordinate (y1) at (0,.5,2);
			\coordinate (y2) at (0,1.5,2);
			\coordinate (y3) at (0,2.5,2);
			\coordinate (y4) at (0,3.5,2);
		\end{scope}
		\draw[-stealth] ($(x0)+(0,-.1)$) -- node[midway,sloped,below] {0-th and 1-st channels} ($(y0)+(0,-.1)$);
		\draw[-stealth] ($(y0)+(0,-.1)$) -- node[midway,sloped,below] {2-nd and 3-rd channels} ($(z0)+(.1,0)$);
		\draw[-stealth] ($(z0)+(.1,0)$) -- node[midway,sloped,below] {4-th channel} ($(z1)+(.1,0)$);
		\node at ($(x1)+(0,.2)$) {$00$};
		\node at ($(x2)+(0,.2)$) {$01$};
		\node at ($(x3)+(0,.2)$) {$11$};
		\node at ($(x4)+(0,.2)$) {$10$};
		\node at ($(y1)+(-.2,.1)$) {$00$};
		\node at ($(y2)+(-.2,.1)$) {$01$};
		\node at ($(y3)+(-.2,.1)$) {$11$};
		\node at ($(y4)+(-.2,.1)$) {$10$};
	\end{tikzpicture}
	\vskip -.8em
	\caption{The RPG of the $(2,2,2,2,2)$-ary codebook %$\{(\epsilon,\epsilon, 1, 0, 0)$, $(0, \epsilon, \epsilon, 1, 0)$, $(0, 0, \epsilon, \epsilon, 1)$, $(1, 0, 0, \epsilon, \epsilon)$, $(\epsilon, 1, 0, 0, \epsilon)\}$
	in \cref{eg:5d}.}
	\label{fig:3d4}
	\end{figure}
	\end{minipage}~~
	\begin{minipage}[t]{.45\textwidth}
	\vskip -1em
	\begin{figure}[H]
%\begin{figure}[b]
	\vskip 1em
	\centering
	\tiny
	\tdplotsetmaincoords{70}{30}
	\begin{tikzpicture}[scale=.6]
		\begin{scope}[tdplot_main_coords]
			\foreach \x in {0,1,2,3,4}
			\foreach \y in {0,1,2}
			\foreach \z in {0,1,2}
			{
				\draw (\x,0,\z) -- (\x,2,\z);
				\draw (0,\y,\z) -- (4,\y,\z);
				\draw (\x,\y,0) -- (\x,\y,2);
			}

			\fill[blue,opacity=.3] (0,1,0) -- ++(2,0,0) -- ++(0,1,0) -- ++(0,0,1) -- ++(-2,0,0) -- ++(0,-1,0) -- cycle;
			\fill[red,opacity=.3] (1,0,0) -- ++(2,0,0) -- ++(0,1,0) -- ++(0,0,1) -- ++(-2,0,0) -- ++(0,-1,0) -- cycle;
			\fill[green,opacity=.3] (0,0,1) -- ++(1,0,0) -- ++(0,2,0) -- ++(0,0,1) -- ++(-1,0,0) -- ++(0,-2,0) -- cycle;
			\fill[gray,opacity=.3] (3,0,0) -- ++(1,0,0) -- ++(0,1,0) -- ++(0,0,2) -- ++(-1,0,0) -- ++(0,-1,0) -- cycle;

			\coordinate (x0) at (0,0,0);
			\coordinate (y0) at (4,0,0);
			\coordinate (z0) at (4,2,0);
			\coordinate (z1) at (4,2,2);

			\coordinate (x1) at (.5,2,2);
			\coordinate (x2) at (1.5,2,2);
			\coordinate (x3) at (2.5,2,2);
			\coordinate (x4) at (3.5,2,2);
		\end{scope}
		\draw[-stealth] ($(x0)+(0,-.1)$) -- node[midway,sloped,below] {0-th and 1-st channels} ($(y0)+(0,-.1)$);
		\draw[-stealth] ($(y0)+(0,-.1)$) -- node[midway,sloped,below] {2-nd channel} ($(z0)+(.1,0)$);
		\draw[-stealth] ($(z0)+(.1,0)$) -- node[midway,sloped,below] {3-rd channel} ($(z1)+(.1,0)$);
		\node at ($(x1)+(0,.2)$) {$00$};
		\node at ($(x2)+(0,.2)$) {$01$};
		\node at ($(x3)+(0,.2)$) {$11$};
		\node at ($(x4)+(0,.2)$) {$10$};
	\end{tikzpicture}
	\vskip -1em
	\caption{A $(2, 2, 2, 2)$-ary selvage core.}
	\label{fig:3d3}
\end{figure}
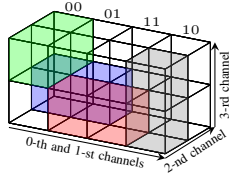
	\end{minipage}
	\end{minipage}
	\ifnum\paperversion=1
	\vskip -1.5em
	\fi
\end{figure}

\begin{example}
Below is the codeword matrix of a $(2,2,2,2)$-ary selvage core with its RPG shown in \cref{fig:3d3}:
\begin{equation*}
	\begin{matrix}
		\text{\tiny red}\\
		\text{\tiny blue}\\
		\text{\tiny green}\\
		\text{\tiny gray}
	\end{matrix}
	\begin{pmatrix}
		\epsilon & 1 & 0 & 0\\
		0 & \epsilon & 1 & 0\\
		0 & 0 & \epsilon & 1\\
		1 & 0 & 0 & \epsilon
	\end{pmatrix}
	\begin{matrix}
		\leftarrow \text{probability } (q_1 q_2 q_3)^{-1},\\
		\leftarrow \text{probability } (q_0 q_2 q_3)^{-1},\\
		\leftarrow \text{probability } (q_0 q_1 q_3)^{-1},\\
		\leftarrow \text{probability } (q_0 q_1 q_2)^{-1}.
	\end{matrix}
\end{equation*}
%In general, the construction works for any $Q = (q_i)_{i \in \ZZ_n}$ with $q_i \geq 2$.
\end{example}

	We can see from the above example that the RPG of a selvage code has a simple interweave structure:
	Each block in the interweave core has exactly one edge fitting an edge of the container in the same dimension.
	The RPGs in \cref{fig:3d1} fall in this category.
	For each plane, this structure looks like a selvage at the edges of denim, so we name our construction ``selvage''.

\begin{table}
	\centering
	\caption{The Descriptive Codeword Lengths of Optimal Codes on SPA}
	\label{tab:entropy}
	\vskip -.6em
	\begin{tabular}{r|c|c}
		\toprule
		& Optimal Prefix (Entropy) & Optimal Tree-Decodable\\ \midrule
		$(2,2,2)$-ary & 1.559581 & 1.559581 \\
		$(5,3,2)$-ary & 2.976887 & 2.980124\\
		$(6,3,2)$-ary & 3.154833 & 3.154833\\
		\bottomrule
	\end{tabular}
	\vskip -2em
\end{table}

After that, we pad sufficiently many codewords of length $\langle 1, \ldots, 1 \rangle$, each assigned with a probability $\prod_{i \in \ZZ_n} q_i^{-1}$, so that the sum of probabilities becomes $1$ while the local redundancy is zero everywhere.
This corresponds to filling the RPG with as many unit hyper-cubes as possible until the container is full.
The fully-filled RPG corresponds to a prefix code as the blocks do not overlap.
Since the local redundancy of each codeword is zero, the overall redundancy of the code is also zero.
In other words, the $Q$-ary selvage code achieves the entropy on the $Q$-ary SPA, which is the multiset of the probabilities assigned above.
On the other hand, since for each channel $i$ there exists a codeword in the interweave core such that its $i$-th component is $\epsilon$, we know from \cref{thm:E} that the selvage code is not tree-decodable.

We now formally state the above construction.

\begin{definition}[Selvage Code]
	Let $n \ge 3$.
	The \emph{$Q$-ary selvage core}, denoted by $\core_Q$, is a codebook $\{\ccore_j\}_{j \in \ZZ_n}$ where for every $j \in \ZZ_n$,
	\begin{equation*}
		\ccore_j(i) = \begin{cases}
			\epsilon & \text{if } i = j,\\
			1 & \text{if } i = (j + 1) \bmod n,\\
			0 & \text{otherwise}.
		\end{cases}
	\end{equation*}
	The \emph{$Q$-ary selvage code}, denoted by $\wc_Q$, is the codebook
	$\core_Q \cup \{ \mathbf{s} \in \prod_{i \in \ZZ_n} \ZZ_{q_i} \colon \mathbf{s} \prefix \mathbf{c}, \forall \mathbf{c} \in \core_Q \}$.
\end{definition}

%Besides stating that the selvage core and selvage code are prefix codes that are not tree decodable, the theorem below also
%%The lemma below
%counts the number of unit hyper-cubes we have introduced.
%It is simply the hyper-volume of the container minus the hyper-volume occupied by the selvage core.

\begin{theorem} \label{thm:interweave}
	Let $n \ge 3$.
	Then, $| \wc_Q \setminus \core_Q | = \prod_{i \in \ZZ_n} q_i - \sum_{i \in \ZZ_n} q_i$.
	Also, both $\core_Q$ and $\wc_Q$ are $Q$-ary non-tree-decodable prefix codes.
\end{theorem}

%The lemma below counts the number of unit hyper-cubes we have introduced.
%It is simply the hyper-volume of the container minus the hyper-volume occupied by the selvage core.

%\begin{lemma} \label{lem:count}
%	Let $n \ge 3$.
%	$| \wc_Q \setminus \core_Q | = \prod_{i \in \ZZ_n} q_i - \sum_{i \in \ZZ_n} q_i$.
%\end{lemma}

\noindent 
Given the number of unit hyper-cubes introduced, we can define the $Q$-ary SPA on which the $Q$-ary selvage code achieves the entropy.

\begin{definition}[SPA] \label{def:spa}
	Let $n \ge 3$ and $k \coloneqq | \wc_Q \setminus \core_Q |$.
	The \emph{$Q$-ary selvage probability assembly (SPA)} is the multiset of probabilities
	\begin{equation*}
		\textstyle
		\left( \biguplus_{j \in \ZZ_n} \left\{q_j \prod_{i \in \ZZ_n} q_i^{-1} \right\} \right)
		\uplus
		\left( \biguplus_{j \in \ZZ_k} \left\{ \prod_{i \in \ZZ_n} q_i^{-1} \right\} \right).
	\end{equation*}
\end{definition}

\begin{theorem} \label{thm:spa_optimal}
	Let $n \ge 3$.
	The $Q$-ary selvage code $\wc_Q$ achieves the entropy on the $Q$-ary SPA.
\end{theorem}

\subsection{Sufficient Condition}

%We prove that if $Q$ is $t$-separable for some $t \geq 3$, then $Q$ is above tree line. As discussed at the beginning of this section, selvage codes and SPAs play major roles in our proof, where we show that any $Q$-ary optimal prefix code on a $Q'$-ary SPA must have the same codeword lengths as those of the $Q'$-ary selvage code.

%Before proving the claim, let us
Before we start, we first
examine the expected descriptive codeword lengths of the (entropy-achieving) $Q$-ary selvage code and an optimal $Q$-ary tree-decodable code on the $Q$-ary SPA for some choices of $Q$.
\Cref{tab:entropy} shows the expected lengths for $Q$ being $(2,2,2)$, $(5,3,2)$, or $(6,3,2)$, where the optimal tree-decodable codes are found using the multichannel Huffman procedure described in~\cite{prefix}.

For the $(2,2,2)$-ary case, we observe that optimal tree-decodable codes are entropy-achieving.
This is as expected since, when all channel alphabets in $Q$ are of the same size $q$, a $Q$-ary code is equivalent to a single-channel $q$-ary code~\cite{yao10}, and single-channel Huffman codes are optimal prefix codes.

%Moving onto the cases with asymmetric channels,
For the $(5,3,2)$-ary case, we observe that all $170,625$ possible trees produced by the multichannel Huffman procedure~\cite{prefix} have non-zero redundancy.
On the other hand, for the $(6,3,2)$-ary case, some of the $1,467,357$ trees achieve the entropy.
This suggests that besides the case where all channels having the same alphabet size, there are other $Q$'s which are below tree line.
%This suggests that all channels having the same alphabet size does not determine whether $Q$ is above or below tree line.
In fact, we know that $Q = (6,3,2)$ is below tree line because each $6$-ary symbol can be split into a tuple consisting of a $2$-ary and a $3$-ary symbol, and all two-channel prefix codes are tree-decodable~\cite{prefix}.

% Now consider the $(5,3,2)$-ary SPA.
% Due to~\cite{prefix}, it suffices to consider adding at most $1$ dummy symbol (with $0$ probability) for these channels to construct a multichannel Huffman code (optimal tree-decodable).
% Then, we test all $170,625$ trees produced by the Huffman procedure and find the one with the minimal redundancy.
% From~\cref{tab:entropy}, the optimal tree-decodable code has a non-zero redundancy.
%
% For the $(6,3,2)$-ary SPA, we need to add at most $2$ dummy symbols for the Huffman code.
% After testing all $1,467,357$ trees, the optimal one achieves the entropy bound.
%
% In other words, besides the case that the alphabet sizes are the same, there are other cases that the SPA does not have a chasm.
% The following theorem shows when there is a chasm, and then we will explain why there is no chasm otherwise.

We now define the notion of $t$-separation which will serve as a sufficient condition for $Q$ being above tree line.
For convenience, we regard the tuple $Q$ as a multiset of the channel alphabets.

\begin{definition}[$t$-Separation]\label{def:separation}
	Let $Q' \subseteq Q$ and write $\bar{Q}' \coloneqq Q \setminus Q'$.
	We say that $Q'$ is \emph{separated} from $Q$ if any non-negative integral solution $(x_q)_{q \in Q} \in \NN^{|Q|}$ to the equation
	\begin{equation}\label{eq:separation}
		%\textstyle
		\prod_{q \in Q} q^{x_q} = \prod_{q \in \bar{Q}'} q
	\end{equation}
	satisfies $x_q = 0$ for all $q \in Q'$.
	Let $\{Q_j\}_{j \in \ZZ_t}$ be a partition of $Q$, i.e., $Q = \biguplus_{j \in \ZZ_t} Q_j$.
	We say that $\{Q_j\}_{j \in \ZZ_t}$ is a \emph{$t$-separation} of $Q$ if, for each $j \in \ZZ_t$, $Q_j$ is separated from $Q$.
	We say that $Q$ is \emph{$t$-separable} if there exists a $t$-separation of $Q$.
\end{definition}

A linear-algebraic interpretation of \cref{def:separation} is to view \cref{eq:separation} as a system of linear Diophantine equations $\mathbf{A}_Q \mathbf{x} = \mathbf{A}_Q \mathbf{e}_{\bar{Q}'}$ defined as follows:
$\mathbf{A}_Q$ is a matrix determined by $Q$ with rows indexed by prime factors of elements in $Q$ and columns indexed by elements in $Q$.
The $(p,q)$-th entry of $\mathbf{A}_Q$, where $p$ is a prime and $q \in Q$, is the exponent of $p$ in the unique prime factorization of $q$.
The vectors $\mathbf{x}$ and $\mathbf{e}_{\bar{Q}'}$ have their entries indexed by elements in $Q$, and $\mathbf{e}_{\bar{Q}'}$ is the binary vector where those entries indexed by $\bar{Q}'$ are $1$ and other entries are $0$.
The separation condition is equivalent to that if $\mathbf{A}_Q \mathbf{x} = \mathbf{A}_Q \mathbf{e}_{\bar{Q}'}$ for %some non-negative integral
$\mathbf{x} \in \NN^{|Q|}$, then the entries of $\mathbf{x}$ indexed by $\bar{Q}'$ must be 0.

\begin{example}
%Using the above linear-algebraic view, one can easily check whether small examples of $Q$ are separable.
%For example,
Consider $Q = (4,6,10,15)$ and hence
\begin{equation*}
	\mathbf{A}_Q =
	\begin{matrix}
		\text{\tiny p=2}\\
		\text{\tiny p=3}\\
		\text{\tiny p=5}
	\end{matrix}
	\begin{pmatrix}
		2 & 1 & 1 & 0 \\
		0 & 1 & 0 & 1 \\
		0 & 0 & 1 & 1
	\end{pmatrix} \stackrel{\text{row op.}}{\sim}
	\mathbf{A}_Q' =
	\begin{pmatrix}
		2 & 2 & 0 & 0 \\
		0 & 1 & 0 & 1 \\
		0 & 0 & 1 & 1
	\end{pmatrix}.
\end{equation*}
We can easily see that $\{4\}$ is not separated from $Q$ because
$$\mathbf{A}_Q
(1, 0, 0, 2)^\intercal
%\begin{pmatrix}
%	1 & 0 & 0 & 2
%\end{pmatrix}^\intercal
=
\mathbf{A}_Q
(0, 1, 1, 1)^\intercal,$$
%\begin{pmatrix}
%	0 & 1 & 1 & 1
%\end{pmatrix}^\intercal.
where $(0, 1, 1, 1)^\intercal = \mathbf{e}_{Q \setminus \{4\}}$.
Similarly, we observe that each of $\{6\}$, $\{10\}$ and $\{15\}$ is also not separated from $Q$.
We therefore conclude that $Q$ has no $3$- nor $4$-separation as each of them must contain a singleton chunk.
On the other hand, by transforming $\mathbf{A}_Q$ to $\mathbf{A}_Q'$ via elementary row operations,
%\[
%	\mathbf{A}_Q \sim
%	\begin{pmatrix}
%		2 & 2 & 0 & 0 \\
%		0 & 1 & 0 & 1 \\
%		0 & 0 & 1 & 1
%	\end{pmatrix}.
%\]
we see that any solution $\mathbf{x} \in \NN^4$ satisfying
$$(2,2,0,0) \mathbf{x}
=
(2,2,0,0) (0,0,1,1)^\intercal$$
must have its first two entries, i.e., those corresponding to $\{4,6\}$, set to $0$.
Similar holds for
$$(0,0,1,1) \mathbf{x}
=
(0,0,1,1) (1,1,0,0)^\intercal,$$ which
concerns the entries of $\mathbf{x}$ corresponding to $\{10,15\}$.
We therefore conclude that $\{\{4,6\}, \{10,15\}\}$ is a $2$-separation of $Q$.
\end{example}

The following lemma gives a natural class of separations of $Q$, which can be used to identify certain special cases with ease.

\begin{lemma}[Natural Separation]
	\label{lem:natural_separation}
	A partition $\{Q_j\}_{j \in \ZZ_t}$ of $Q$ is a $t$-separation of $Q$ if, for each $j \in \ZZ_t$ and for every $q' \in Q_j$, there exists a prime $p_{q'}$ such that
	$p_{q'} \mid q'$ but $p_{q'} \nmid q$ for all $q \in \biguplus_{k \in \ZZ_t \setminus \{j\}} Q_k$.
	%A partition $\{Q_j\}_{j \in \ZZ_t}$ of $Q$ is a $t$-separation of $Q$ if, for each $j \in \ZZ_t$, there exists a prime $p_j$ such that
	%$p_j \mid q$ for all $q \in Q_j$, but $p_j \nmid q$ for all $q \in \biguplus_{k \in \ZZ_t \setminus \{j\}} Q_k$.
	%$p_j$ divides all $q \in Q_j$ but does not divide any $q \in \bar{Q}_j$, where $\bar{Q}_j \coloneqq \biguplus_{k \in \ZZ_t \setminus \{j\}} Q_k$.
\end{lemma}

%\begin{IEEEproof}
%	Fix any $j \in \ZZ_t$.
%	Let $(x_q)_{q \in Q} \in \NN^{|Q|}$ solves \cref{eq:separation} where $Q' = Q_j$.
%	%%We claim that $x_q = 0$ for all $q \in Q_j$ by considering the two cases which $p_j$ satisfy.
%	%%Indeed, observe that $p_j$ divides $q$ for all $q \in Q_j$ and $p_j$ does not divide $q$ for any $q \in \bar{Q}_j$.
%	If $x_{q'} \neq 0$ for some $q' \in Q_j$, then
%	$p_{q'} \mid \prod_{q \in Q} q^{x_q}$ but $p_{q'} \nmid \prod_{q \in \bar{Q}'} q$, which contradicts that $(x_q)_{q \in Q}$ solves \cref{eq:separation}.
%	%If $x_q \neq 0$ for some $q \in Q_j$, then
%	%$p_j \mid \prod_{q \in Q} q^{x_q}$ but $p_j \nmid \prod_{q \in \bar{Q}'} q$, which contradicts that $(x_q)_{q \in Q}$ solves \cref{eq:separation}.
%	%%$p_j$ divides the LHS of \cref{eq:separation} but not the RHS, which is a contradiction.
%\end{IEEEproof}

If for each $j \in \ZZ_t$, there exists a prime $p_j$ such that $p_j \mid q$ for all $q \in Q_j$ but $p_j \nmid q$ for all $q \in \biguplus_{k \in \ZZ_t \setminus \{j\}} Q_k$, we can also apply \cref{lem:natural_separation} to conclude the $t$-separation of $Q$.
This lemma also
%\Cref{lem:natural_separation}
covers the $n$-separation where $Q = (q_i)_{i \in \ZZ_n}$ is a tuple of distinct primes. %\footnote{%
%	By slightly modifying the proof of this lemma, we can have a more general natural separation statement:
%	A partition $\{Q_j\}_{j \in \ZZ_t}$ of $Q$ is a $t$-separation of $Q$ if, for each $j \in \ZZ_t$ and for every $q' \in Q_j$, there exists a prime $p_{q'}$ such that
%	$p_{q'} \mid q'$ but $p_{q'} \nmid q$ for all $q \in \biguplus_{k \in \ZZ_t \setminus \{j\}} Q_k$.
%}
For such $Q$, a natural $n$-separation consists of $Q_i = \{q_i\}$ for all $i \in \ZZ_n$.
There are, however, separations of $Q$ which do not satisfy the condition of~\cref{lem:natural_separation}, e.g.,
the separation $\{\{6\}, \{4\}, \{3\}\}$ of $Q = (6,4,3)$. %\footnote{%
	A reason is that the power of $q$ on the RHS of \cref{eq:separation} is fixed to $1$, which means that we do not require the set of columns of $\mathbf{A}_Q$ to be a basis.
%}

\begin{theorem}\label{thm:main}
	Let $t \geq 3$. If $Q$ is $t$-separable, then $Q$ is above tree line.
\end{theorem}

We remark that $Q$ being $t$-separable does not imply that any superset $Q' \supset Q$ is $t$-separable. Indeed, we can check that $(4, 6, 10)$ is $3$-separable but $(4, 6, 10, 15)$ is not.
This aligns with our geometric intuition that, when there are more dimensions, it is easier to pack blocks into a container without them interweaving with each other.

%In the proof of \cref{thm:main}, we showed that if $Q$ is $t$-separable then $Q$ is above tree line on the $Q^\times$-ary SPA. 
The proof of \cref{thm:main} can be done by showing that if $Q$ is $t$-separable, then $Q$ is above tree line on the $Q^\times$-ary SPA where $Q^\times := (\prod_{q \in Q_j} q)_{j \in \ZZ_t}$ is the product channel alphabets.
In the following, we state a partial converse -- if $Q$ is above tree line on the $Q^\times$-ary SPA then $Q$ is $t$-separable. We note, however, that this is not the converse of \cref{thm:main} since, for $Q$ to be above tree line, it could be above tree line for some $P$ different from the $Q^\times$-ary SPA.

\begin{theorem} \label{thm:converse}
	Let $\{Q_j\}_{j \in \ZZ_t}$ be a $t$-partition of $Q$ for some $t \ge 3$.
	If $Q$ is above tree line on the $Q^\times$-ary SPA,
	where $Q^\times = (\prod_{q \in Q_j} q)_{j \in \ZZ_t}$,
	then $\{Q_j\}_{j \in \ZZ_t}$ is a $t$-separation.
\end{theorem}

We outline the high level idea of the proof of \cref{thm:converse}.
Suppose $Q_{j^*}$ is not separated from $Q$, then $\prod_{q \in Q} q^{x_q} = \prod_{\bar{Q}_j} q$ admits a solution $(x^*_q)_{q \in Q}$ where $x^*_q \neq 0$ for some $q \in Q_{j^*}$.
We show how to construct a $Q$-ary tree-decodable code which is entropy-achieving on the $Q^\times$-ary SPA, contradicting the assumption.
The basic idea is to first interpret the $Q^\times$-ary selvage code as a $Q$-ary code, then use the above equation to move codeword symbols in the selvage core across different channels without affecting the descriptive length of each codeword.
The second step ``disentangles'' the selvage core, making it tree-decodable.

\ifnum\paperversion=2
More concretely, suppose $Q^+$ denotes the set of $q$ for which $x^*_q > 0$.
We consider two cases: 1) $q_{i^*} > 2$ for some $q_{i^*} \in Q_{j^*} \cap Q^+$; 2) $q = 2$ for all $q \in Q_{j^*} \cap Q^+$.

For Case 1, we change the $j^*$-th codeword in the selvage code so that its $i^*$-th component is %$\mathbf{2}^{x^*_{q_{i^*}}}$, 
an all-$2$ string of length $x^*_{q_{i^*}}$,
its $i$-th component is 
%$\mathbf{0}^{x^*_{q_{i}}}$ 
an all-$0$ string of length $x^*_{q_{i}}$
for all $i \in Q^+ \setminus \{q_{i^*}\}$, and all other components are empty strings.
Consequently, the $i^*$-th components of all codewords are all non-empty, which allows us to build a decoding tree with a root assigned to channel $i^*$.

For Case 2, we can derive that all $q \in \bar{Q}_{j^*} \setminus Q^+$ are powers of $2$. Therefore we can write $q_{i^\dagger} = q_{i^*}^r$ for some $q_{i^\dagger} \in \bar{Q}_{j^*} \setminus Q^+$ and some $q_{i^*} \in Q_{j^*} \cap Q^+$.
Using this relation, we can replace each $q_{i^\dagger}$-ary symbol in the selvage code with $r$ $q_{i^*}$-ary symbol (while choosing symbols carefully so that prefix-freeness is preserved). 
Consequently, channel $i^\dagger$ becomes a dummy channel and the $i^*$-th component of all codewords are non-empty.
The latter again allows us to build a decoding tree with a root assigned to channel $i^*$.
\fi

\section{Concluding Remarks}

We investigated the interweave structure of non-tree-decodable prefix codes and then filled a theoretical gap by giving a sufficient condition for the existence of optimal multichannel tree-decodable codes that are not optimal prefix codes.
We leave proving that (a relaxation of) the sufficient condition is necessary as one of the future research directions on the theory of multichannel source coding.

\ifnum\paperversion=2
\appendices

\section{Proof of \cref{thm:guillotine}}

\begin{IEEEproof}
If a prefix code is not tree decodable, then it means that there is a subspace $S$ guillotine-cut from the container, or $S$ is the container itself, such that $S$ is occupied by more than one block but no guillotine-cut is possible without cutting through a block.
That is, there is no way to obtain any block in $S$ by guillotine-cuts.

Conversely, suppose some blocks cannot be obtained by guillotine-cuts.
We consider a subset $T$ of these blocks having the same prefix where $|T| > 1$ such that the smallest space $S$ containing the blocks in $T$ is not possible to be separated by guillotine-cuts without cutting through a block.
Suppose the prefix code is tree decodable, then there is a non-leaf node corresponds to the space $S$.
However, there is no way to further guillotine-cut $S$, so we cannot find a class for this node, which contradicts that the code is tree decodable.
\end{IEEEproof}

\section{Proof of \cref{thm:E}}

\begin{IEEEproof}
	Suppose $M$ is tree decodable.
	The root node of the decoding tree must belong to one of the classes in $\ZZ_n$.
	However, for each possible class $k$, there exists at least one codeword that is not a descendant of the root node.
	This contradicts that $M$ is tree decodable.
\end{IEEEproof}

\section{Proof of \cref{thm:interweave}}

\begin{IEEEproof}
	Recall that, for every channel $i \in \ZZ_n$ and every codeword $j \in \ZZ_n$ in the core, we have
	\begin{equation*}
		\ccore_j(i) = \begin{cases}
			\epsilon & \text{if } i = j,\\
			1 & \text{if } i = (j + 1) \bmod n,\\
			0 & \text{otherwise}.
		\end{cases}
	\end{equation*}

	For any $j,k \in \ZZ_n$ with $k \notin \{j, (j+1) \bmod n\}$, we have
	$\ccore_j((j+1) \bmod n) = 1$ and $\ccore_k((j+1) \bmod n) = 0$, which means
	$\ccore_j \prefix \ccore_k$ in the $((j+1) \bmod n)$-th channel.
	On the other hand, for $k = (j+1) \bmod n$, we have $\ccore_j((k+1) \bmod n) = 0$ and $\ccore_k((k+1) \bmod n) = 1$, which means $\ccore_j \prefix \ccore_k$ in the $((k+1) \bmod n)$-th channel.
	Summarizing, we have $\ccore_j \prefix \ccore_k$ for all $j, k \in \ZZ_n$ with $j \neq k$, i.e., $\core_Q$ is a prefix code.
	Let $M$ be the codeword matrix of $\core_Q$ where the $j$-th row is $\ccore_j$.
	It is clear that $\mathcal{E}_M(i) = \{i\}$ for all $i \in \ZZ_n$, so $\core_Q$ is not tree decodable by \cref{thm:E}.

	Now, we consider $\wc_Q$.
	Note that the words in $\prod_{i \in \ZZ_n} \ZZ_{q_i}$ are distinct but have the same length tuple $\langle 1 \rangle_{i \in \ZZ_n}$, so every pair of words in this set are prefix-free.
	As $\core_Q$ is a prefix code, $\wc_Q$ is also a prefix code by its definition.
	Let $M'$ be the codeword matrix of $\wc_Q$ where the first $t$ rows are the matrix $M$.
	Again, we have $\mathcal{E}_{M'}(i) = \{i\}$, so by \cref{thm:E}, $\wc_Q$ is also not tree decodable.
%\end{IEEEproof}

%\section{Proof of \cref{lem:count}}

We now calculate the size of $|\wc_Q \setminus \core_Q|$.
%\begin{IEEEproof}
	%By \cref{thm:interweave},
	Recall that $\wc_Q$ is a prefix code.
	In this code, $\ellmax_i = 1$ for all $i \in \ZZ_n$, so the container in its RPG is a hyper-rectangle of size $\langle q_i \rangle_{i \in \ZZ_n}$.
	The codeword $\ccore_j \in \core_Q$ corresponds to a block of size $\langle q_i^{1-\delta_{i,j}} \rangle_{i \in \ZZ_n}$, where $\delta_{i,j}$ is the Kronecker delta, so $\vol(\ccore_j) = q_j$.
	The smallest possible block has size $\langle 1 \rangle_{i \in \ZZ_n}$, which corresponds to a codeword having the same length as any of those in $\wc_Q \setminus \core_Q$.
	That is, $\vol(\mathbf{c}) = 1$ for any $\mathbf{c} \in \wc_Q \setminus \core_Q$.
	As a prefix code, the blocks for $\wc_Q$ must packable in the container.
	The construction of $\wc_Q$ includes all codewords of length $\langle 1 \rangle_{i \in \ZZ_n}$ which are prefix-free to those in $\core_Q$, which means that besides the blocks for $\core_Q$, we put as many blocks of size $\langle 1 \rangle_{i \in \ZZ_n}$ as possible to fully fill the container.
	In other words, the sum of volumes of the blocks equals the volume of the container, i.e.,
	\begin{equation*}
		\sum_{i \in \ZZ_n} \vol(\ccore_i) + \smashoperator{\sum_{\mathbf{c} \in \wc_Q \setminus \core_Q}} \vol(\mathbf{c}) = \prod_{i \in \ZZ_n} q_i.
	\end{equation*}
	The proof is done by reordering the terms.
\end{IEEEproof}

\section{Proof of \cref{thm:spa_optimal}}

\begin{IEEEproof}
	We define the source code as follows.
	For each $j \in \ZZ_n$, we map the probability $q_j \prod_{i \in \ZZ_n} q_i^{-1}$ to $\ccore_j \in \core_Q$.
	We have
	\begin{equation} \label{eq:ipc_eq1}
		\left|\ccore_j\right| = \sum_{i \in \ZZ_n} \ln q_i - \ln q_j = -\ln \left( q_j \prod_{i \in \ZZ_n} q_i^{-1} \right).
	\end{equation}
	Next, we map each of the remaining probabilities $\prod_{i \in \ZZ_n} q_i^{-1}$ to an arbitrary codeword $\mathbf{c} \in \wc_Q \setminus \core_Q$ bijectively, which is possible according to \cref{thm:interweave}. %\cref{lem:count}.
	Recall that $\len(\mathbf{c}) = \langle 1 \rangle_{i \in \ZZ_n}$, so we have
	\begin{equation} \label{eq:ipc_eq2}
		|\mathbf{c}| = \sum_{i \in \ZZ_n} \ln q_i = -\ln \prod_{i \in \ZZ_n} q_i^{-1}.
	\end{equation}
	\cref{eq:ipc_eq1,eq:ipc_eq2} imply that the condition for the equality of the entropy bound holds.
\end{IEEEproof}

\section{Proof of \cref{lem:natural_separation}}

\begin{IEEEproof}
	Fix any $j \in \ZZ_t$.
	Let $(x_q)_{q \in Q} \in \NN^{|Q|}$ solves \cref{eq:separation} where $Q' = Q_j$.
	%%We claim that $x_q = 0$ for all $q \in Q_j$ by considering the two cases which $p_j$ satisfy.
	%%Indeed, observe that $p_j$ divides $q$ for all $q \in Q_j$ and $p_j$ does not divide $q$ for any $q \in \bar{Q}_j$.
	If $x_{q'} \neq 0$ for some $q' \in Q_j$, then
	$p_{q'} \mid \prod_{q \in Q} q^{x_q}$ but $p_{q'} \nmid \prod_{q \in \bar{Q}'} q$, which contradicts that $(x_q)_{q \in Q}$ solves \cref{eq:separation}.
	%If $x_q \neq 0$ for some $q \in Q_j$, then
	%$p_j \mid \prod_{q \in Q} q^{x_q}$ but $p_j \nmid \prod_{q \in \bar{Q}'} q$, which contradicts that $(x_q)_{q \in Q}$ solves \cref{eq:separation}.
	%%$p_j$ divides the LHS of \cref{eq:separation} but not the RHS, which is a contradiction.
\end{IEEEproof}

\section{Proof of \cref{thm:main}}

\begin{IEEEproof}
	Let $\{Q_j\}_{j \in \ZZ_t}$ be a $t$-separation of $Q$, $Q^\times = (q^\times_i)_{i \in \ZZ_t}$ be the product channel alphabets where $q^\times_i = \prod_{q \in Q_i} q$ for $i \in \ZZ_t$, and $P$ be the $Q^\times$-ary SPA.
	By~\cref{thm:spa_optimal}, the $Q^\times$-ary selvage code $\wc_{Q^\times}$ on $P$ achieves the entropy bound.
	Interpreting $\wc_{Q^\times}$ as a $Q$-ary code, we obtain an entropy-achieving $Q$-ary prefix code on $P$.

	%To show that there is no entropy-achieving $Q$-ary tree-decodable code on $P$
	%It thus remains to show that no $Q$-ary tree-decodable code achieves the entropy bound on $P$.

	Suppose there is an entropy-achieving $Q$-ary tree-decodable code $C$ on $P$.
	%Suppose towards a contradiction that there exists a $Q$-ary tree-decodable code $C$ which achieves the entropy bound on $P$.
	Recall that %the $Q^\times$-ary SPA $P$ contains the probabilities $q^\times_j \prod_{i \in \ZZ_t} (q^\times_i)^{-1}$ for all $j \in \ZZ_t$.
	$q^\times_j / \prod_{i \in \ZZ_t} q^\times_i \in P$ for all $j \in \ZZ_t$.
	Let $\langle x_{i,j} \rangle_{i \in \ZZ_n}$ be the length of the codeword for the probability $q^\times_j / \prod_{i \in \ZZ_t} q^\times_i$.
	Since $C$ achieves the entropy bound, we have
	%$|\langle x_{i,j} \rangle_{i \in \ZZ_n}|_Q = -\ln (q_j'/\prod_{i \in \ZZ_t} q_i')$, which gives
	$\prod_{i \in \ZZ_n} q_i^{x_{i,j}} = (\prod_{i \in \ZZ_t} q^\times_i) / q^\times_j = \prod_{q \in Q \setminus Q_j} q$.
%	\[
%		\prod_{i \in \ZZ_t} q_i^{x_{i,j}} = (q^\times_j)^{-1} \prod_{i \in \ZZ_t} q^\times_i = \prod_{q \in \bar{Q}_j} q.
%	\]
	Since $\{Q_j\}_{j \in \ZZ_t}$ is a $t$-separation of $Q$, for each $j \in \ZZ_t$, we must have $x_{i,j} = 0$ for all $q_i \in Q_j$. %and for all $j \in \ZZ_t$.
	In other words, for every channel $i \in \ZZ_n$, there exists a codeword in $C$ such that its $i$-th component is $\epsilon$.
	By~\cref{thm:E}, $C$ is not tree-decodable, which is a contradiction.
\end{IEEEproof}

\section{Proof of \cref{thm:converse}}

\begin{IEEEproof}
	In the proof below, we adopt the following notion.
	For any strings $a, b$, denote by $a \| b$ the concatenation of $a$ and $b$. %, e.g., $01 \| 10$ means the string $0110$.
	To represent a string formed by duplicating the same symbol, we bold the symbol and write the number of repetitions as its exponent. %, e.g., $\mathbf{0}^4$ means the string $0000$.
	We will always use $i$ and $j$ as running variables over $\ZZ_n$ and $\ZZ_t$ respectively.

	We prove \cref{thm:converse} by contrapositive.
	Suppose for some $j^* \in \ZZ_t$, $Q_j$ is not separated from $Q$, i.e., there is some $(x^*_q)_{q \in Q} \in \NN^{|Q|}$ solving
	\cref{eq:separation}
	such that $x^*_q \neq 0$ for some $q \in Q_{j^*}$.
	We construct an entropy-achieving $Q$-ary tree-decodable code on the $Q^\times$-ary SPA.

	Since $(x^*_q)_{q \in Q}$ solves \cref{eq:separation}, we have
	\begin{align*}
		\prod_{q \in Q} q^{x^*_q} &= \prod_{q \in \bar{Q}_{j^*}} q.
	\end{align*}
	Define $Q^+ \coloneqq \{q \in Q \colon x^*_q > 0\} \subseteq Q$.
	We have
	\begin{equation}\label{eq:case1_len}
		\prod_{q \in Q^+} q^{x^*_q} = \prod_{q \in \bar{Q}_{j^*}} q.
	\end{equation}
	Note also that $Q_{j^*} \cap Q^+ \neq \emptyset$ and $\bar{Q}_{j^*} \setminus Q^+ \neq \emptyset$.

	For $j \in \ZZ_t$, let $q_j^\times = \prod_{q \in Q_j} q$ be the product channel alphabet sizes.
	By \cref{thm:spa_optimal}, the $Q^\times$-ary selvage code $\wc$ is entropy-achieving on the $Q^\times$-ary SPA.
	The $Q^\times$-ary selvage core $\core$ can be interpreted as a $Q$-ary code using the following transform:
	For each $Q^\times$-ary codeword $\mathbf{c}^\times$ of $\wc$, define the $Q$-ary codeword $\mathbf{c}$ by $\mathbf{c}(i) = \mathbf{c}^\times(j)$ where $q_i \in Q_j$ for all $i \in \ZZ_n$ and $j \in \ZZ_t$.
	This is possible as $\mathbf{c}^\times(j) \in \{\epsilon, 0, 1\}$.
	We denote this $Q$-ary form of the selvage core by $\overline{\core}$.
	It is easy to check that $\overline{\core}$ is a prefix code whose codewords have one-to-one correspondence to those of $\core$, and each codeword in $\overline{\core}$ has the same descriptive length as that of its counterpart in $\core$.

	Case 1: $q_{i^*} > 2$ for some $q_{i^*} \in Q_{j^*} \cap Q^+$.
	We modify the core $\overline{\core} = \{\overline{\ccore_j}\}_{j \in \ZZ_t}$ to construct a new codebook $\widehat{\core} = \{\widehat{\ccore_j}\}_{j \in \ZZ_t}$ by $\widehat{\ccore_j} = \overline{\ccore_j}$ for $j \neq j^*$, and
	\begin{equation*}
		\widehat{\ccore_{j^*}}(i) = \begin{cases}
			\mathbf{2}^{x^*_{q_i}} & \text{if } i = i^*, \\
			\mathbf{0}^{x^*_{q_i}} & \text{if } q_i \in Q^+ \setminus \{q_{i^*}\}, \\
			\epsilon & \text{otherwise}.
		\end{cases}
	\end{equation*}
	By \cref{eq:case1_len}, we know that $|\widehat{\ccore_{j^*}}| = |\overline{\ccore_{j^*}}|$.

	As $\overline{\core}$ is a prefix code, $\overline{\core} \setminus \{\overline{\ccore_{j^*}}\}$ also is.
	From the $i^*$-th component, we know that $\widehat{\ccore_{j^*}} \prefix \widehat{\ccore_j}$ for all $j \in \ZZ_t \setminus \{j^*\}$.
	Therefore, $\widehat{\core}$ is a prefix code.

	Next, we extend the core $\widehat{\core}$ into a prefix code $\widehat{C}$ by building its decoding tree:
	\begin{enumerate}[1)]
		\item The root node belongs to class $i^*$.
		\item The $0$-child of the root node is the root of a subtree constructed by first building a tree whose leaves are labelled by the set $\{0\} \times \prod_{q \in Q \setminus \{q_{i^*}\}} \ZZ_q$, and then removing the subtree under the node labelled by $\widehat{\ccore_j}$ for all $j \in \ZZ_t \setminus \{j^*, (j^*-1) \bmod t\}$. Let $S_0$ denote the set of leaves of this subtree except $\widehat{\ccore_j}$ for all $j \in \ZZ_t \setminus \{j^*, (j^*-1) \bmod t\}$.
		\item The $1$-child of the root node is the root of a subtree constructed by first building a tree whose leaves are labelled by the set $\{1\} \times \prod_{q \in Q \setminus \{q_{i^*}\}} \ZZ_q$, and then removing the subtree under the node labelled by $\widehat{\ccore_{(j^*-1) \bmod t}}$. Let $S_1$ denote the set of leaves of this subtree except $\widehat{\ccore_{(j^*-1) \bmod t}}$.
		\item The $2$-child of the root node is the root of a subtree constructed by first building a tree whose leaves are labelled by the set $\{2\} \times \ZZ_{q_{i^*}}^{x^*_{q_{i^*}} - 1} \times \prod_{q \in Q^+ \setminus \{q_{i^*}\}} \ZZ_q^{x^*_q} \times \prod_{q \in Q_j} \ZZ_q$, and then removing the subtree under the node labelled by $\widehat{\ccore_{j^*}}$. Let $S_2$ denote the set of leaves of this subtree except $\widehat{\ccore_{j^*}}$.
		\item For $i = 3, \ldots, q_{i^*} - 1$, the $i$-child of the root node is the root of a subtree whose leaves are labelled by the set $\{i\} \times \prod_{q \in Q \setminus \{q_{i^*}\}} \ZZ_q$. Let $S_3$ denote the set of leaves of these subtrees.
	\end{enumerate}

	We now count $|\widehat{C} \setminus \widehat{\core}| = \sum_{i \in \ZZ_4} |S_i|$.
	We observe that
	\begin{IEEEeqnarray*}{rCl}
		|S_0| & = & \prod_{i \in \ZZ_n \setminus \{i^*\}} q_i - \sum_{j \in \ZZ_t \setminus \{j^*, (j^*-1) \bmod t\}} q_j^\times,\\
		|S_1| & = & \prod_{i \in \ZZ_n \setminus \{i^*\}} q_i - q_{(j^*-1) \bmod t}^\times,\\
		|S_2| & = & q_{i^*}^{x^*_{q_{i^*}}-1} \prod_{q \in Q^+ \setminus \{q_{i^*}\}} q^{x^*_q} \prod_{q \in Q_j} q - q_{j^*}^\times \\
		& = & q_{i^*}^{-1} \prod_{q \in \bar{Q}_{j^*}} q \prod_{q \in Q_j} q - q_{j^*}^\times \\
		& = & \prod_{i \in \ZZ_n \setminus \{i^*\}} q_i - q_{j^*}^\times, \\
		|S_3| & = & (q_{i^*} - 3) \prod_{i \in \ZZ_n \setminus \{i^*\}} q_i.
	\end{IEEEeqnarray*}
	Summing them up, we have $|\widehat{C} \setminus \widehat{\core}| = \prod_{i \in \ZZ_n} q_i - \sum_{j \in \ZZ_t} q_j^\times$, which is the same size as $|\wc_Q \setminus \core_Q| = \prod_{j \in \ZZ_t} q_j^\times - \sum_{j \in \ZZ_t} q_j^\times$ according to \cref{thm:interweave}.
	Therefore, we have a bijective mapping from $\wc_Q$ to $\widehat{C}$ where the descriptive length of every codeword is preserved.
	That is, $\widehat{C}$ is an entropy-achieving $Q$-ary tree-decodable code on the $Q^\times$-ary SPA.

	Case 2: $q = 2$ for all $q \in Q_{j^*} \cap Q^+$.
	We can see from \cref{eq:case1_len} that every $q \in \bar{Q}_{j^*} \setminus Q^+$ is a power of $2$.
	Pick any $i^*, i^\dagger \in \ZZ_n$ and $j^\dagger \in \ZZ_t$ such that $q_{i^*} \in Q_{j^*} \cap Q^+$ and $q_{i^\dagger} \in \bar{Q}_{j^*} \setminus Q^+$.
	We have
	\begin{equation} \label{eq:case2_len}
		q_{i^\dagger} = q_{i^*}^r
	\end{equation}
	for some $r \in \ZZ^+$.
	We modify the core $\overline{\core} = \{\overline{\ccore_j}\}_{j \in \ZZ_t}$ to construct a new codebook $\widehat{\core} = \{\widehat{\ccore_j}\}_{j \in \ZZ_t}$ by setting $\widehat{\ccore_{j^\dagger}} = \overline{\ccore_{j^\dagger}}$,
	\begin{equation*}
		\widehat{\ccore_{(j^\dagger-2) \bmod t}}(i) =
		\begin{cases}
			\overline{\ccore_j}(i) \| \mathbf{1}^r & \text{if } i = i^*,\\
			\epsilon & \text{if } i = i^\dagger, \\
			\overline{\ccore_j}(i) & \text{if } i \in \ZZ_n \setminus \{i^*,i^\dagger\}
		\end{cases}
	\end{equation*}
	and
	\begin{equation*}
		\widehat{\ccore_j}(i) =
		\begin{cases}
			\overline{\ccore_j}(i) \| \mathbf{0}^r & \text{if } i = i^*,\\
			\epsilon & \text{if } i = i^\dagger, \\
			\overline{\ccore_j}(i) & \text{if } i \in \ZZ_n \setminus \{i^*,i^\dagger\}
		\end{cases}
	\end{equation*}
	for $j \in \ZZ_t \setminus \{j^*,j^\dagger\}$.
	Note that $|\widehat{\ccore_j}| = |\overline{\ccore_j}|$ for all $j \in \ZZ_t$ due to \cref{eq:case2_len}.
	Moreover, we have $\widehat{\ccore_j}(i^\dagger) = \epsilon$ for all $j \in \ZZ_t$ by construction, i.e., channel $i^\dagger$ becomes dummy.

	Except for $(j,j') = ((j^\dagger-2) \bmod t, (j^\dagger-1) \bmod t)$, for all other distinct $j, j' \in \ZZ_t$, the original codewords $\overline{\ccore_j}$ and $\overline{\ccore_{j'}}$ are prefix-free in some channel $i \neq i^\dagger$, so do the modified codewords $\widehat{\ccore_j}$ and $\widehat{\ccore_{j'}}$. For $(j,j') = ((j^\dagger-2) \bmod t, (j^\dagger-1) \bmod t)$, we note that $\widehat{\ccore_j}(i^*) = \overline{\ccore_j}(i^*) \| \mathbf{1}^r$ and $\widehat{\ccore_{j'}}(i^*) = \overline{\ccore_{j'}}(i^*) \| \mathbf{0}^r$. There are four possibilities: $(\overline{\ccore_j}(i^*),\overline{\ccore_{j'}}(i^*)) \in \{(\epsilon, 0), (1,\epsilon), (0,1), (0,0)\}$. Clearly, in each case, $\widehat{\ccore_j}(i^*)$ and $\widehat{\ccore_{j'}}(i^*)$ are prefix-free. We conclude that $\widehat{\core}$ is a prefix code.

	Next, we extend the core $\widehat{\core}$ into a prefix code $\widehat{C}$ by building its decoding tree.
	The tree is constructed by first building a tree whose leaves are labelled by the set $\ZZ_{q_{i^*}}^{1+r} \times \ZZ_{q_{i^\dagger}}^0 \times \prod_{i \in \ZZ_n \setminus \{i^*,i^\dagger\}} \ZZ_{q_i}$, and then removing the subtrees under the nodes labelled by $\widehat{\ccore_j}$, for $j \in \ZZ_t$.

	Observe that
	\begin{equation*}
		|\widehat{C} \setminus \widehat{\core}| = q_{i^*}^{1+r} \prod_{i \in \ZZ_n \setminus \{i^*,i^\dagger\}} q_i - \sum_{i \in \ZZ_t} q_i^\times = \prod_{i \in \ZZ_n} q_i - \sum_{i \in \ZZ_t} q_i^\times.
	\end{equation*}
	By the same argument as in Case 1, $\widehat{C}$ is an entropy-achieving $Q$-ary tree-decodable code on the $Q^\times$-ary SPA.
\end{IEEEproof}

\fi

\balance
\bibliographystyle{IEEEtran}
\bibliography{prefix2-bib}

\end{document}